\documentclass[%
 sor4-1,
 reprint,
 amsmath,amssymb,
]{revtex4-1}

\usepackage{graphicx}
\usepackage{dcolumn}
\usepackage{bm}
\usepackage{xcolor}
\usepackage{enumerate}

\begin{document}
	
\preprint{APS/123-QED}

\title{A Data-Driven Method for Automated Data Superposition \\ with Applications in Soft Matter Science}
\author{Kyle R. Lennon$^1$}
\thanks{Electronic mail: krlennon@mit.edu}
\author{Gareth H. McKinley$^2$}
\thanks{Electronic mail: gareth@mit.edu}
\author{James W. Swan$^1$}
\thanks{Deceased}
\affiliation{1. Department of Chemical Engineering, \\ 2. Department of Mechanical Engineering, Massachusetts Institute of Technology, Cambridge, MA \\}
\date{\today}

\begin{abstract}
The superposition of data sets with internal parametric self-similarity is a longstanding and widespread technique for the analysis of many types of experimental data across the physical sciences. Typically, this superposition is performed manually, or recently by one of a few automated algorithms. However, these methods are often heuristic in nature, are prone to user bias via manual data shifting or parameterization, and lack a native framework for handling uncertainty in both the data and the resulting model of the superposed data. In this work, we develop a data-driven, non-parametric method for superposing experimental data with arbitrary coordinate transformations, which employs Gaussian process regression to learn statistical models that describe the data, and then uses maximum \emph{a posteriori} estimation to optimally superpose the data sets. This statistical framework is robust to experimental noise, and automatically produces uncertainty estimates for the learned coordinate transformations. Moreover, it is distinguished from black-box machine learning in its interpretability -- specifically, it produces a model that may itself be interrogated to gain insight into the system under study. We demonstrate these salient features of our method through its application to four representative data sets characterizing the mechanics of soft materials. In every case, our method replicates results obtained using other approaches, but with reduced bias and the addition of uncertainty estimates. This method enables a standardized, statistical treatment of self-similar data across many fields, producing interpretable data-driven models that may inform applications such as materials classification, design, and discovery.

\end{abstract}

\keywords{---}

\maketitle

\section{Introduction}

For many of the physical processes encountered in scientific and engineering studies, the underlying governing equations are unknown, oftentimes despite tremendous scientific effort. Even in systems for which it is possible to write down the governing equations, the dynamical response often remains too complicated to admit a compact analytical solution, and frequently even numerical solutions to such complicated equation sets are too cumbersome to form the basis of useful tools for the prediction or analysis of experimental observations. In such cases, insights and predictions are instead often made by using domain knowledge and heuristics to identify patterns in experimental data. These patterns may then inform empirical models that describe the behavior of systems quite accurately, over a wide range of conditions. With increasing popularity, these patterns and empirical models are being identified automatically by computational algorithms, a process commonly referred to as \emph{machine learning} \cite{Carleo2019,Butler2018,Ferguson2017}.

There are many ways in which patterns in data may inform useful mathematical models without any physical considerations. Here we consider one special case, which we exemplify by examining a simple physical system whose governing equations, and the analytical solution to those equations, are known: one-dimensional diffusion from an instantaneous point source. In this system, a fixed mass $M$ of a passive scalar is released at the origin ($x = 0$) at time $t = 0$ in an infinite domain, and allowed to diffuse with a constant diffusion coefficient $D$, such that the concentration $C$ of the diffusing species is governed by:
\begin{equation}
    \frac{\partial C}{\partial t} = D\frac{\partial^2 C}{\partial x^2}.
\end{equation}
The analytical solution to this diffusion problem for all times $t > 0$ is known:
\begin{equation}
    C(x,t) = \frac{M}{\sqrt{4\pi D t}}\exp\left(-\frac{x^2}{4 D t}\right).
\end{equation}

\begin{figure*}
    \centering
    \includegraphics[width=\textwidth]{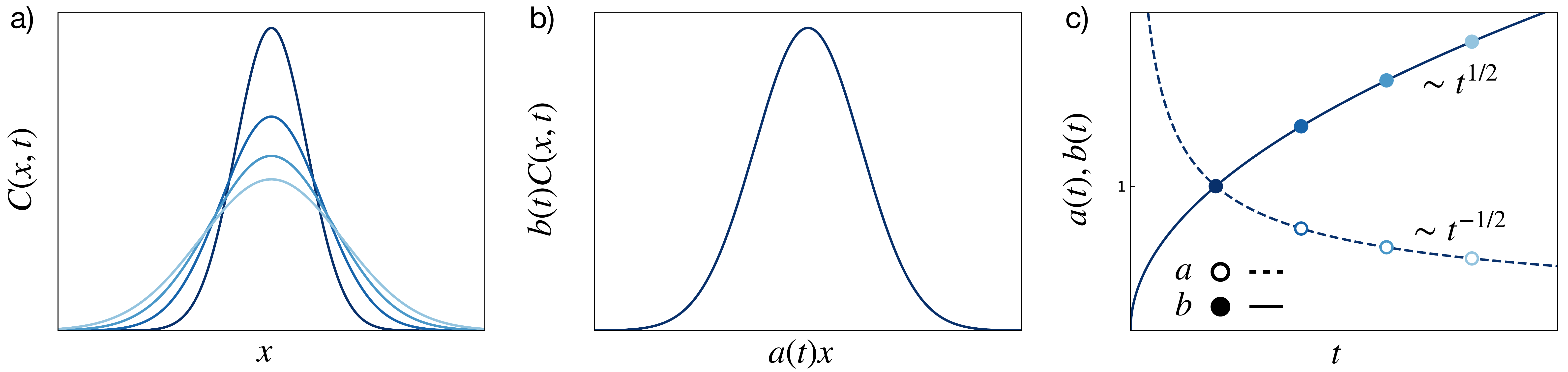}
    \caption{Example of fictitious measurements of the concentration profile of a diffusing species from an instantaneous point source. (a) The concentration profile measured instantaneously at four different times, with lighter shaded curves representing later times. (b) A `master curve' constructed by rescaling the width and height of the concentration profiles by time-dependent `shift factors' $a(t)$ and $b(t)$, respectively. (c) The shift factors $a(t)$ and $b(t)$ plotted as a function of time. The earliest-time concentration profile is taken as the reference, so its shift factors are unity. The remaining shift factors exhibit the trends $a(t) \sim t^{-1/2}$ and $b(t) \sim t^{1/2}$.}
    \label{fig:diffusion}
\end{figure*}

However, consider if the underlying diffusion process and analytical solution for this system were \emph{not} known. Instead, we have access to a (possibly noisy) device that measures the concentration profile $C(x,t)$ at an instant in time. Figure \ref{fig:diffusion}a presents some of these fictional measurements. A trained researcher may notice that the concentration profiles at each measured time form a set of self-similar curves, each with a different width and maximum value. Noting this pattern, they might rescale the height of each concentration by a factor $b(t)$, and rescale the width by a factor $a(t)$, so that all the rescaled curves fall on top of one another. This superposition creates the so-called `master curve' shown in Figure \ref{fig:diffusion}b. The researcher records the horizontal and vertical `shift factors', $a(t)$ and $b(t)$, respectively, and plots them as a function of time, noting that $a(t) \sim t^{-1/2}$ and $b(t) \sim t^{1/2}$. Therefore, using ideas of self-similarity \cite{Barenblatt2003} they propose the following model for the concentration profile:
\begin{equation}
    C(x,t) = \frac{A}{\sqrt{t}} g\left(\frac{x}{\sqrt{t}}\right).
\end{equation}
We recognize that, for this problem, the functional form for the master curve $g(z)$ is a simple Gaussian in the reduced variable $z \equiv x/\sqrt{t}$.  An approximation for this form might be determined from the rescaled data through the same sort of methods that determined forms for $a(t)$ and $b(t)$, at which point a self-similar solution to the diffusion equation will have been constructed directly from data.

In nonlinear or higher-dimensional systems, it may not be possible to choose simple functional forms that describe a master curve or the shift factors $a(t)$ and $b(t)$. However, parametric self-similarity in data -- that is, the observation that we may construct a master curve by transforming discrete data observed at different `states' as illustrated in the diffusion example above -- is a widespread and salient principle in the analysis of thermophysical property data in many fields, including rheology, solid mechanics, and material science \cite{leaderman1944, tobolsky1945, ferry1980}. In fact, the same principle of dynamical self-similarity finds applications in many different fields as well, such as electromagnetism and even finance \cite{wagner1915, lillo2003}. 

Self-similar data sets may be employed to make accurate predictions of material characteristics at intermediate states, or to learn physical features about the underlying dynamics or structure of the system under study \cite{Barenblatt2003,deGennes1979}, often times by traditional methods such as direct interpolation between data or by constitutive modeling. However, as demonstrated by the previous example, there is another method for analyzing self-similar data. This method is most accurately referred to as the `method of reduced variables', but is commonly referred to as `data superposition' or the development of `master curves' \cite{leaderman1944,ferry1980}. When applicable, this method is potent, as it is agnostic to selection of an underlying constitutive model, and may encompass a broader and physically robust set of hypotheses than direct interpolation between specific (possibly noisy) data. Thus, when new cases of data self-similarity are explored, data superposition can be immediately applied to gain physical insight and develop predictive tools, without the need to develop new models.

In many cases, including the diffusion example discussed above, an experiment measuring a property with parametric self-similarity $ C( x; t ) $, over an independent variable $ x $ and with a state parameter $ t $, is related to a master curve $ g( z ) $ in the reduced variable $z \equiv a(t)x$ by the similarity relation:
\begin{equation}
    C( x; t ) = \frac{1}{b(t)} g\left( a(t) x \right) + c\, h( x ). \label{eq:master_curve}
\end{equation}
Here $ a( t ) $ and $ b( t ) $ are state-parameter--dependent shift factors, while $ h( x ) $ and $c$ are a state-parameter--independent offset function and multiplicative constant, respectively. Together, these transformations collapse the experimental measurements onto the master curve (denoted generically $g(z)$) for all values of $ x $. Most often, the master curve is constructed from superposition of noisy experimental data.  This process uses experimental data for $ C( x; t ) $ taken for a finite set of state parameters $ \{ t_j \} $ and over a fixed range of $ x $.  Then one determines the set of shift factors, $ \{ a_j = a( t_j ) \} $, $ \{ b_j = b( t_j ) \} $, and $ c $ so that these finite data sets across experiments are brought into registry without advance knowledge of the functional form of the master curve $ g( z ) $.

Typically the process of master curve construction is done by an expert in the art who determines the shift factors, $ \{ a_j \} $, $ \{ b_j \} $ and $ c $, by eye.  This can be done reliably, but it is a time-consuming process and may not be robust to experimental uncertainties.  If one aims to construct master curves from large sets of data in an unbiased fashion -- where bias could come in the form of presuming an existing model for any of $ g( z ) $, $ a( t ) $ or $ b( t ) $, or through the preconceptions and biases of the expert analyzing the data -- then an algorithm for the automated construction of such master curves is necessary.  In this work, we present such an algorithm relying on Gaussian process regression and maximum \textit{a posteriori} estimation and then apply the algorithm to examples using rheological data for soft material systems drawn from the existing literature.

At the start of this Introduction, we noted that the automatic identification of an empirical model, such as a master curve, from self-similar data represents a form of machine learning. Indeed, in the sense that the algorithm presented herein takes as input a set of data and outputs a trained model (here, a master curve and set of shift factors) to make predictions at states outside of the training set, it represents an instance of \emph{supervised learning} -- a class of machine learning algorithms seeking to map inputs to outputs based on a set of known input-output pairs \cite{Carleo2019}. At the same time, the goal of the method of reduced variables is often to obtain the master curve and shift factors themselves, which may be interrogated further to study the physics of the material system, independent of any specific model. In this case, the training data set does not include any examples of the desired output (i.e. the master curve and shift factors); rather, the algorithm is tasked with identifying patterns within the data -- an instance of \emph{unsupervised learning} \cite{Carleo2019}. This mode of learning is reminiscent of principal component analysis (PCA), a popular unsupervised learning strategy in soft materials science that also uses patterns in data sets to reduce their dimensionality \cite{Carleo2019,Wagner2016,Alghooneh2019}.

Although the method presented in this work may be classified as a machine learning algorithm, it is worth briefly distinguishing it from a black-box machine learning approach \cite{Rudin2019}. A black-box approach would use the experimental data set, with labels $(x, t)$ and targets $C(x; t)$, to train an empirical model $f(x, t)$ mapping labels to predicted targets, for example in the form of an artificial neural network \cite{Yegnanarayana2009}. This empirical model, however, would be uninterpretable outside of the predictions that it makes, meaning that nothing about the underlying system may be learned by examining the form of $f(x, t)$ directly. Our approach, on the other hand, represents a form of \emph{interpretable} machine learning \cite{Molnar2020,Molnar2020book}, where the learned model itself may provide insight to the mechanisms governing the system's behavior. For instance, the shape of the master curves or the functional form of the inferred $a(t)$ and $b(t)$ curves may be connected to physical processes (e.g. Fickian diffusion in our example) in real systems. In studies of physical systems, the development of interpretable models such as ours is essential, in that they not only provide predictive tools, but may be examined to discover physical explanations that underpin their specific predictions \cite{brenner2021}.

\subsection{Examples in Soft Matter Science}

Many thermomechanical properties of interest in soft materials satisfy principles of self-similarity in the time domain or with respect to rate of deformation on variation of an appropriate state parameter \cite{plazek1965,larsen2008,dekker2018,caggioni2020,gupta2012,struik1977,lalwani2021}, which may be attributed to the wide range of temporal and spatial scales that govern the underlying dynamics \cite{markovitz1975,deGennes1979,Barenblatt2003}. Some examples include self-similar curves for:
\begin{itemize}
    \item the creep compliance as a function time on variation of the temperature in polymer melts (\emph{time-temperature superposition}) \cite{plazek1965},
    \item the linear viscoelastic modulus as a function of frequency on variation of extent of reaction during gelation (\emph{time-cure superposition}) \cite{larsen2008}, 
    \item the shear stress as a function of shear rate on variation of packing fraction in an emulsion (\emph{shear rate-volume fraction superposition}) \cite{dekker2018},
    \item the relaxation modulus as a function of time on variation of the wait time $t_w$ since sample preparation in an aging clay suspension (\emph{time--age-time superposition}) \cite{gupta2012}.
\end{itemize}
It is often challenging to design rheometric experiments and equipment capable of measuring these evolving self-similar mechanical properties over the entire dynamic parameter range on which they vary.  Instead, the principles of self-similarity \cite{Barenblatt2003} and dynamic scaling \cite{deGennes1979}, coupled with a judicious choice of physiochemical state parameters such as temperature or pH are used to expand the effective range of time scales accessible in experiments. 

Furthermore, the construction of master curves and the interpretation of those curves plays a critical role in extrapolating from direct measurement of material properties across a limited number of experiments to fundamental knowledge of the microscale physical processes giving rise to these material properties. For instance:
\begin{itemize}
    \item in time-temperature superposition $C(x; t)$ would be the linear creep compliance (or the relaxation modulus), $ x $ would be the temporal coordinate, $ t $ would be the temperature in a given set of experiments, $ a( t ) $ and $ b( t ) $ would be the horizontal and vertical shift factors that might be used to learn about the temperature dependence of relaxation processes in the soft material. In such a representation, $ c $ would be a fluidity that is independent of the temperature, and $ h( x ) = x $ would indicate linear creep at long times \cite{plazek1965},
    \item in the time-cure superposition principle that underpins gelation \cite{Adolf1990}, $ C( x; t ) $ would be the complex modulus, $ x $ would be the frequency, $ t $ would be perhaps the concentration of gelator, $ a( t ) $ and $ b( t ) $ would be horizontal and vertical shift factors that might be used to learn about the mechanism of gelation, $ c $ would be an offset accounting for the solvent viscosity, and $ h( x ) = i x $ would reflect this high frequency viscous mode \cite{larsen2008},
    \item in the shear rate-volume fraction superposition of an emulsion, $ C( x; t ) $ would be the steady shear stress, $ x $ would be the imposed shear rate, $ t $ would be the packing fraction ($\phi$) of the emulsion, $ a( t ) $ and $ b( t ) $ would be the horizontal and vertical shift factors that might be used to learn about the packing fraction dependence of the yield stress, $ c $ would be a high shear viscosity and $ h( x ) = x $ would reflect Newtonian flow of the emulsion at very high shear rates \cite{dekker2018,caggioni2020},
    \item in time--age-time superposition of an aging clay suspension, $ C( x; t ) $ would be the linear relaxation modulus, $ x $ would be the effective or material time coordinate ($\tilde{t}$), $ t $ would be the wait time ($t_w$) after mixing or pre-shear of the aging suspension, $ a( t ) $ and $ b( t ) $ would be the horizontal and vertical shift factors that might be used to learn about the rates of microstructural yielding in the suspension, $ c $ would be an elastic plateau modulus and $ h( x ) = 1 $ would reflect the non-aging elastic energy stored in the microstructure \cite{gupta2012,struik1977}.
\end{itemize}
Note that in this last case, the independent variable $x$ represents an \emph{effective}, or dynamically rescaled, time coordinate, $x = \xi(t;\mu)$, representing a nonlinear transformation of the laboratory time coordinate $t$ with some scaling exponent $\mu$ \cite{Joshi2018}. Despite this added complexity, the method that we present in this work is sufficiently general to accommodate nonlinear coordinate transformations, and can indeed infer optimal values of the scaling parameters defining these transformations (such as the aging exponent $\mu$) at the same time as the shift factors $ a( t ) $ and $ b( t ) $.

The present work is organized as follows. In Section \ref{sec:method}, we develop the mathematics behind the algorithm for automatic construction of master curves. This includes a brief description of Gaussian process regression, a development of maximum likelihood and maximum \emph{a posteriori} estimation, and the discussion of Monte Carlo cross-validation for hyperparameter optimization. Then, Section \ref{sec:examples} applies this method to the four specific soft materials science examples listed in this Section, which are of increasing complexity. Finally, Section \ref{sec:predictions} demonstrates how the learned master curves, combined with the inferred shift factors $a(t)$ and $b(t)$, may be used to inform forward predictions of data with automatic uncertainty estimates.

\section{Method}
\label{sec:method}

A number of previous works have made efforts to automate the process of master curve construction. These include methods based on minimizing the mean squared error between data at every state and a single basis expansion \cite{honerkamp1993, buttlar1998, sihn1999}, methods for minimizing the area between linear interpolants defined by data at different states \cite{barbero2004, gergesova2011}, methods for minimizing the mean squared error between the derivatives of spline interpolants fit to the data \cite{hermida1994, naya2013}, methods to minimize the arc length between data sets at different states \cite{cho2009, maiti2016}, and a method that leverages mathematical constraints on the Fourier transform of real-valued time-series data \cite{Rouleau2013}. However, only a few of these methods have been demonstrated for simultaneous horizontal and vertical shifting, and many require either parameters, an appropriate interpolant, or a set of basis functions to be specified by the user, thereby introducing elements of subjectivity into the methodology. Moreover, few of these methods explicitly account for the fact that experimental data possesses a finite amount of noise and measurement uncertainty, which may affect both the resulting master curve and confidence in the inferred shift factors. Because noise is not treated directly by these approaches, systematic parametric sensitivity analysis has required computationally intensive approaches, such as bootstrap resampling \cite{maiti2019}, which may require hundreds of thousands of evaluations of the full superposition algorithm.

Here, we propose an approach to automated master curve construction that is both data-driven and statistically motivated, thereby enabling rapid bidirectional superposition of data with automatic uncertainty quantification. We rely on Gaussian process regression (GPR) to infer a continuous and probabilistic description of the underlying data, and apply maximum \textit{a posteriori} estimation to infer the best set of parameters that shift the noisy data onto a common master curve. This methodology is not limited to any specific form for the master curve, nor is it constrained to any specific form for the transformations applied to the data to obtain the master curve. Thus, it may be applied to cases of simple horizontal shifting, simultaneous horizontal and vertical shifting, as well as nonlinear coordinate transformations such as material time dilation in rheologically aging systems. Moreover, this approach is non-parametric, meaning that users need only supply data and the functional form of the transformations that should be applied to obtain the master curve. Finally, because we use a probabilistic description of the data and statistical comparisons of different data sets, the underlying noise in the data and the resulting uncertainties in the inferred shift factors are handled naturally within this framework.

In the remainder of this section, we will develop our approach, beginning with a brief discussion of Gaussian process regression, then continuing to discussions of inference. We develop our maximum \textit{a posteriori} approach by first considering the slightly simpler formalism of maximum likelihood estimation. We subsequently incorporate prior expectations about the shifting parameters $a(t)$ and $b(t)$ into the framework in the form of prior distributions, which turns the maximum likelihood estimates of the shift parameters to maximum \textit{a posteriori} estimates.

\subsection{Gaussian Process Regression}

Gaussian process regression (GPR) -- known in some fields as \emph{kriging} \cite{matheron1963} -- is a machine learning method for obtaining a probabilistic model of a stochastic process from data \cite{rasmussen2006}. GPR assumes that the data $C(x;t)$ at fixed $t$ are described by a Gaussian process (GP):
\begin{equation}
    C(x;t) \sim GP(\mu(x;t), K(\theta, x, x'; t)), \label{eq:gp}
\end{equation}
with a mean function $\mu(x; t)$ and a covariance function, commonly referred to as the kernel of the GP, $K(\theta, x, x'; t)$, between any two points $x$ and $x'$, with a set of hyperparameters $\theta$. In this Section, we use the symbol `$\sim$' to signify that a variable (or series of variables) behaves according to some statistical distribution (or stochastic process). Informally, a GP represents a distribution in \emph{function space}, or an ensemble of functions most likely to describe the data set. The form of the \emph{prior} mean and covariance functions, $\mu(x; t)$ and $K(\theta, x, x'; t)$ encode prior expectations about the functions in this distribution, such as noise structure, smoothness, and variations over $x$.

With prior expectations over fitting functions specified by the form of $\mu(x; t)$ and $K(\theta, x, x'; t)$, GPR proceeds by storing data points in the kernel matrix $\boldsymbol{K}$ with $K_{ij} = K(\theta, x_i, x_j; y)$, where $x_i$ and $x_j$ represent values of $x$ present in the input data set, and determining the hyperparameters $\theta$ that maximize the log-marginal likelihood of observing the data from the GP model \cite{rasmussen2006}. The GP model now represents a distribution over functions that fit the data set subject to prior expectations, with \emph{posterior} mean $m(x^*; t)$ and variance $s(x^*; t)^2$ at an unmeasured point $x^*$ describing a Gaussian distribution $\mathcal{N}$ over the predicted value, $C(x^*; t) \sim \mathcal{N}(m(x^*; t), s(x^*; t)^2)$. For a Gaussian process with zero-mean ($\mu(x; t) = 0$), these posterior mean and variance functions are:
\begin{equation}
    m(x^*; t) = \underline{K}(\theta, x^*, \underline{x}; t)\boldsymbol{K}^{-1}\underline{C}(\underline{x}; t),
\end{equation}
\begin{equation}
    s(x^*)^2 = K(\theta, x^*, x^*; t) - \underline{K}(\theta, x^*, \underline{x};t)\boldsymbol{K}^{-1}\boldsymbol{K}^T, \nonumber
\end{equation}
where $\underline{x}$ represents the vector of all values of $x$ in the training data set, $\underline{C}(\underline{x}; t)$ represents the vector with elements $C_i = C(x_i; t)$ for all $x_i$ in the training set, and $\underline{K}(\theta, x^*, \underline{x}; t)$ represents the vector with elements $K_i = K(\theta, x^*, x_i; t)$ for all $x_i$ in the training set.

In this work, we assume a zero-mean GP and a combination of constant, white noise, and rational quadratic kernels \cite{rasmussen2006}:
\begin{align}
    & K(\theta, x, x'; t) = \\
    & A(t)\left(1 + \frac{(x - x')^2}{2\alpha(t) l(t)^2}\right)^{-\alpha(t)} + B(t) + (\sigma(t)^2 + \sigma_u^2) \delta_{x, x'}, \nonumber
\end{align}
where $\theta = [A(t), B(t), \alpha(t), l(t), \sigma(t)]$ are the hyperparameters which may depend on the state parameter $t$, and $\delta_{x,x'}$ is the Kronecker delta function between points $x$ and $x'$. Notably, the form of this kernel allows for the GP to automatically adapt to the level of noise in the data set, via the hyperparameter $\sigma(t)$. The state-independent hyperparameter $\sigma_u$ represents an experimental uncertainty level which is not fit by GPR, but rather can be specified in advance to encode additional experimental uncertainty that is not evident in variations of the data \cite{ewoldt2015, singh2019}. Other authors have noted that a reasonable limit for the relative uncertainty level in high-quality rheological data is 4\% \cite{freund2015}; therefore, unless otherwise stated, we take $\sigma_u = 0.04$ when the GP model describes the logarithm of experimental data. This value may be assumed by any user to keep the method non-parametric; however, if the relative uncertainty of specific classes of experimental data is known in advance, then $\sigma_u$ may be adjusted by the user.

We will not discuss the details of the fitting protocol for GPR or hyperparameter optimization here, but refer interested readers to the following references for more information on GP models and GPR: \cite{rasmussen2006, gortler2019, duvenaud2014}.

\subsection{Maximum Likelihood Estimation}

\subsubsection{Linearly Scaled Multiplicative Shifting}

The first step in our automated superposition algorithm is to fit each data set at distinct state $t$ to its own GP model. Once this regression is complete, our goal is to optimally superpose these GP models to create a master curve. Because GPs are probabilistic in nature, it is sensible to define a statistical criterion for the optimal superposition.

To begin in developing this criterion, consider the problem of registering a single data point, $(x_i, C_i)_j$ from a data set at state $t_j$ with a GP model trained on a data set at a different state, $t_k$. For simplicity, consider $t_k$ as the reference state, such that all shift factors required to collapse the data onto the master curve are applied relative to this state. We apply a horizontal shift $a_{jk}$ and vertical shift $b_{jk}$ to the data point in $t_j$, bringing it to the coordinates $(a_{jk} x_i, b_{jk} C_i)$. This shifted data point is now treated as an observation drawn from the GP model at state $t_k$:
\begin{equation}
    b_{jk} C_i \sim \mathcal{N}(m_k(a_{jk} x_i), s_k(a_{jk} x_i)^2).
\end{equation}
The likelihood $p_k(x_i, C_i | a_{jk}, b_{jk})$ of this observation may be evaluated from the Gaussian probability density function taken from the GP at the coordinate $a_{jk} x_i$. It will be convenient to work with the negative of the logarithm of this likelihood (abbreviated NLL, for negative log-likelihood):
\begin{align}
    - \ln p_k(x_i, C_i | a_{jk}, b_{jk}) =& \,\frac{1}{2}\left(\frac{b_{jk} C_i - m_k(a_{jk} x_i)}{s_k(a_{jk} x_i)}\right)^2 \label{eq:nll} \\
    & + \ln s_k(a_{jk} x_i) + \ln \sqrt{2\pi}. \nonumber
\end{align}
The NLL defines a \emph{loss function} for the superposition of the single data point $(x_i, C_i)_j$ at state $t_j$ and the entire data set at state $t_k$, which is captured by the GP model with mean and variance functions $m_k(x)$ and $s_k(x)^2$, respectively. The likelihood is conditioned on particular values of the shift factors $a_{jk}$ and $b_{jk}$ applied to data at state $t_j$; thus, minimizing the NLL gives the maximum likelihood estimates of $a_{jk}$ and $b_{jk}$ for this particular data point.

Before we extend this approach to include an entire data set, it is worthwhile to consider what exactly the maximum likelihood estimate aims to achieve. We see that the first term in equation \ref{eq:nll} is minimized when $b_{jk} C_i = m_k(a_{jk} x_i)$ -- that is, when the data point at $t_j$ is shifted directly onto the mean of the GP at $t_k$. This matches our intuitive sense of optimal superposition. However, this is not the only term in the NLL that is affected by the shift parameters. The second term in equation \ref{eq:nll} penalizes shifting the data point to coordinates at which the GP has large variance. This term will become particularly important when we shift entire data sets, as it prioritizes superposing two data sets in the region with the least uncertainty in their GP models. Because the covariance in our selected GP kernel depends on the Euclidean distance between an unmeasured coordinate $x^*$ and the measured coordinates $x'$, uncertainty in the GP model is typically greatest in regions where data are sparse, or in regions outside of the range of measured data. Penalizing superposition in these regions relative to regions in which data are dense provides a natural means for regularizing excessive extrapolation.

Equation \ref{eq:nll} develops the loss function for a single data point, based on maximum likelihood estimation. This loss may be applied independently to each data point $(x_i, C_i)_j$ in the data set at state $t_j$, which we denote $\mathcal{D}_j$. We are now interested in the likelihood of observing this entire data set from the GP model at state $t_k$, equivalent to the joint likelihood of observing each data point from that GP. If we treat the data points in $\mathcal{D}_j$ as independent observations, then this joint likelihood is simply the product of the individual likelihoods. The joint NLL is therefore the sum of the individual NLLs:
\begin{equation}
    - \ln p_{jk}(\mathcal{D}_j | a_{jk}, b_{jk}) = \sum_{(x_i, C_i) \in \mathcal{D}_j} - \ln p_k(x_i, C_i | a_{jk}, b_{jk}) \label{eq:joint_nll}
\end{equation}

This equation now defines a loss function over an entire data set, rather than a single data point. It is not yet the final loss function to consider for obtaining a maximum likelihood estimate of the shift parameters, however. Equation \ref{eq:joint_nll} describes the likelihood of observing the data set $\mathcal{D}_j$ to align with the GP model at state $t_k$, but as we shift $\mathcal{D}_j$, we are similarly bringing the data set at the state $t_k$, denoted $\mathcal{D}_k$, close to the GP model trained at state $t_j$. The final loss function should reflect the symmetric nature of this shifting. Thus, we may define the counterparts to equations \ref{eq:nll} and \ref{eq:joint_nll}:
\begin{align}
    - \ln p'_j(x_i, C_i | a_{jk}, b_{jk}) &= \frac{1}{2}\left(\frac{C_i/b_{jk} - m_j(x_i/a_{jk})}{s_j(x_i/a_{jk})}\right)^2 \nonumber \\
    & + \ln s_j(x_i/a_{jk}) + \ln \sqrt{2\pi},
\end{align}
\begin{equation}
    - \ln p'_{jk}(\mathcal{D}_k | a_{jk}, b_{jk}) = \sum_{(x_i, C_i) \in \mathcal{D}_k} - \ln p'_j(x_i, C_i | a_{jk}, b_{jk}). \label{eq:joint_nll_2}
\end{equation}
The shift parameters $a_{jk}$ and $b_{jk}$ now divide $x_i$ and $C_i$, to bring them from state $t_k$ to state $t_j$, and the mean and variance functions $m_j(x^*)$ and $s_j(x^*)$ are now taken from the GP trained at $t_j$.

To complete the loss function, we simply take the joint likelihoods over data sets $\mathcal{D}_j$ and $\mathcal{D}_k$, again assuming independence, such that the likelihood function that we observe both data sets from the same master curve given a shifting parameters $a_j$ and $b_j$ is:
\begin{align}
    - \ln p(\mathcal{D}_j, \mathcal{D}_k | a_{jk}, b_{jk}) =& - \ln p_{jk}(\mathcal{D}_j | a_{jk}, b_{jk}) \label{eq:total_nll}  \\ 
    & - \ln p'_{jk}(\mathcal{D}_k | a_{jk}, b_{jk}). \nonumber
\end{align}
Minimizing this joint NLL loss leads to the maximum likelihood estimates $\hat{a}_{jk}$ and $\hat{b}_{jk}$, the shifting parameters for state $t_j$ relative to state $t_k$:
\begin{equation}
    \{\hat{a}_{jk}, \hat{b}_{jk}\} = \underset{a_{jk},b_{jk}}{\arg \min} \left[- \ln p(\mathcal{D}_j, \mathcal{D}_k | a_{jk}, b_{jk})\right]. \label{eq:max_likelihood_objective}
\end{equation}

When developing a master curve from more than one data set, it is most computationally efficient to perform maximum likelihood estimation pairwise, developing a set of relative shift factors $\{a_{jk}\}$ and $\{b_{jk}\}$ between two curves with consecutive states: $t_j$ and $t_k = t_{j+1}$. Once all relative shift factors have been computed, a global reference state may be selected, and the product or quotient of certain shift factors taken to obtain the global shifts. However, it is straightforward to extend the NLL loss in equation \ref{eq:total_nll} to include more than two data sets, by taking the sum of every pairwise NLL possible between two data sets in the master set. This treatment requires optimization over all shifts $\{a_{jk}\}$ and $\{b_{jk}\}$ simultaneously, a high-dimensional global optimization problem that quickly becomes computationally intensive. Fortunately, pairwise shifting is typically sufficient due to the continuous nature of physical variables; thus, we limit our treatment to pairwise shifting in this work. In the coming sections, we will demonstrate how this approach may be used to solve more complex shifting problems in an efficient manner.

\subsubsection{Logarithmically Scaled Multiplicative Shifting}
\label{sec:log_shifting}

The problem formulation in this section has assumed that the data $(x_i, C_i)$ represent linear-scaled coordinates, and that the desired shifting protocols involve multiplying these coordinates by factors $a_{jk}$ and $b_{jk}$ at each state pair: $t_{j}$, $ t_{k} $. This multiplicative rescaling of data is the most common in many fields of the physical sciences. However, because of the wide underlying spectrum of time and length scales representing the dynamics of soft multiscale materials, rheological data is often represented on logarithmic scales, with the $x$-coordinate representing a temporal coordinate (or its inverse such as frequency or shear rate) that spans orders of magnitude, and with the response coordinate $C$ (commonly a stress or modulus) also spanning many orders of magnitude as well. In this case, we may replace the coordinates $x_i$ and $C_i$ in equations \ref{eq:gp} through \ref{eq:total_nll} with their logarithms, $\ln x_i$ and $\ln C_i$. That is, the GP models are now regressed to the logarithm of the data, and the joint NLL loss is computed based on the logarithm of the data. In this case, the expressions for the rescaled data now involve addition of the logarithm of the shift factor, rather than direct multiplication by the shift factor:
\begin{equation}
    x_i a_{jk} \rightarrow \ln x_i + \ln a_{jk}, \quad x_i/a_{jk} \rightarrow \ln x_i - \ln a_{jk}, \label{eq:a_lin_to_log}
\end{equation}
\begin{equation}
    C_i b_{jk} \rightarrow \ln C_i + \ln b_{jk}, \quad C_i/b_{jk} \rightarrow \ln C_i - \ln b_{jk}.
\end{equation}

This substitution is especially convenient in the case of the vertical shift factor, because it reduces the joint NLL in equation \ref{eq:total_nll} to a quadratic expression in $\ln b_{jk}$. At a fixed value of the lateral shift $\ln a_{jk}$, optimization of the vertical shift factor $\ln b_{jk}$ is now a convex problem whose global minimum can be found analytically. In fact, the minimizer $\ln \hat{b}_{jk}(a_{jk})$ as a function of $a_{jk}$ is given by:
\begin{equation}
    \frac{1}{\sigma^2} = \sum_{x_i \in \mathcal{D}_j}\frac{1}{s_k(\ln x_i + \ln a_{jk})^2} + \sum_{x_i \in \mathcal{D}_k}\frac{1}{s_j(\ln x_i - \ln a_{jk})^2},
\end{equation}
\begin{align}
    & \ln \hat{b}_{jk}(a_{jk}) =  \label{eq:b_minimizer} \\
    & \quad\quad \sigma^2\left[\sum_{(x_i,C_i) \in \mathcal{D}_j}\left(\frac{\ln C_i - m_k(\ln x_i + \ln a_{jk})}{s_k(\ln x_i + \ln a_{jk})^2}\right) \right. \nonumber \\
    & \quad\quad\quad\quad \left. + \sum_{(x_i,C_i) \in \mathcal{D}_k}\left(\frac{\ln C_i - m_j(\ln x_i - \ln a_{jk})}{s_j(\ln x_i - \ln a_{jk})^2}\right)\right]. \nonumber
\end{align}
Note that, since the GP models are trained on the logarithm of the original data, $m_j$ and $s_j$ predict the mean and standard deviation of the logarithm of the material response for state $t_j$ (and similarly for $t_k$). This minimizer function for $\ln b_{jk}$ can be interpreted intuitively as choosing the weighted average of the differences between the data and the opposing GP model, where each difference is weighted inversely by its relative contribution to the total uncertainty in the GP predictions. Also note that the means by which we apply the horizontal shifting do not affect the method for analytically computing the minimizer for the vertical shifts; therefore, we may use equation \ref{eq:b_minimizer} with any type of horizontal shifting, as long as the vertical shift is multiplicative and performed on the logarithm of the data.

The realization that the maximum likelihood estimate for the vertical shift factor $b_{jk}$ may be determined analytically for a specified horizontal shift $a_{jk}$ is significant, as it reduces the pairwise shifting optimization problem from a 2D, non-convex problem to two separate 1D optimization problems, one of which is convex and solved analytically (vertical shifting), and one that is non-convex (horizontal shifting). Whereas the 2D problem requires global optimization techniques such as simulated annealing, which may require large runtimes to thoroughly explore the 2D parameter space, the 1D non-convex problem for optimizing the horizontal shift can be solved by simpler means, such as a grid search and subsequent gradient-based optimization, in a fraction of the time.

\subsubsection{Non-Multiplicative Shifting}
\label{sec:non_multiplicative}

In some circumstances, developing a master curve requires performing transformations that are not simple multiplication by constant scaling factors $a_{jk}$ and $b_{jk}$. In rheology, this may include, for example, subtracting out state-independent viscous contributions from steady flow curve data, $\sigma - \eta\dot{\gamma}$ \cite{Plazek1996}, or a nonlinear `dilation' in the laboratory time coordinate due to power-law rheological aging, $t^{1-\mu} - t_{\mathrm{ref}}^{1-\mu}$ \cite{gupta2012,Joshi2018}. Nonlinear rescaling of temporal and spatial coordinates is also common in the construction of similarity solutions in fluid dynamics and transport phenomena \cite{Barenblatt2003,Eggers2015}. In many such cases, the non-multiplicative shifts are accompanied by subsequent multiplicative shifts. It is possible to generalize the maximum likelihood approach discussed previously to accommodate these more complicated data transformations, with the result again being a high-dimensional, non-convex optimization problem, which is amenable to global optimization techniques such as simulated annealing. However, in many such cases it is possible to leverage the computationally efficient technique developed in the previous section for multiplicative shifting to solve these more complex problems in less time.

To make use of the previous results for non-multiplicative shifting, we may separate the data transformations into a set of non-multiplicative transformations (such as subtracting a state-independent viscous mode, or applying state-independent time dilation) and a set of multiplicative transformations. Typically, there is a single non-multiplicative transformation, which involves a single parameter (such as $\eta$ or $\mu$ in the previously described examples), which we call generically $\varphi$. We may therefore separate the optimization over $\varphi$ and over all multiplicative shift factors $\{a_{jk}\}$ and $\{b_{jk}\}$ into an outer-loop optimization over $\varphi$, and an inner-loop optimization over $\{a_{jk}\}$ and $\{b_{jk}\}$. This technique is reminiscent of hyperparameter optimization, which is typical in many statistical and machine learning approaches, including Gaussian process regression \cite{gelman2013}. In the outer loop, we may perform a grid search over a specified range of $\varphi$, and in the inner loop we perform maximum-likelihood estimation as described in the previous sections for a specific value of $\varphi$, storing the sum of the NLL loss over each pair of data sets for each $\varphi$. Thus, we obtain minimizers $\{\hat{a}_{jk}(\varphi)\}$ and $\{\hat{b}_{jk}(\varphi)\}$ as a function of $\varphi$, and then choose the optimal $\hat{\varphi}$ to minimize the summed NLL loss over all pairwise shifts in the inner loop. This approach is effective due to the efficiency of the inner-loop optimization. Moreover, with a grid search over $\varphi$, the outer-loop optimization is trivially parallelizable; thus it is possible to incorporate non-multiplicative shifts with minimal increase in runtime.

\subsection{Maximum \textit{A Posteriori} Estimation}
\label{sec:posterior}

The maximum likelihood estimation approach described so far represents a systematic and robust method for superposing data, which naturally incorporates our intuitive sense of the role of uncertainty in the superposition process. This approach is sufficient to obtain estimates of the optimal shifting parameters in many, if not most, cases. However, the maximum likelihood approach does not provide estimates of the uncertainty in the inferred shift factors, and it is not effective in the limited cases where the NLL loss is degenerate over multiple values of the shift factor. Both of these deficiencies can be addressed by introducing the notion of a \emph{prior distribution} over the shifting parameters, turning the maximum likelihood estimation approach to one of maximum \textit{a posteriori} estimation.

From Bayes' theorem, the likelihood of observing two data sets from a master curve given certain shift factors, $p(\mathcal{D}_j, \mathcal{D}_k| a_{jk}, b_{jk})$, when multiplied by a prior distribution over $a_{jk}$ and $b_{jk}$, $p(a_{jk}, b_{jk})$, is transformed to the \emph{posterior distribution} over the shifting parameters:
\begin{equation}
    p(a_{jk}, b_{jk} | \mathcal{D}_j, \mathcal{D}_k) \propto p(\mathcal{D}_j, \mathcal{D}_k| a_{jk}, b_{jk})p(a_{jk}, b_{jk}). \label{eq:posterior}
\end{equation}
This transformation brings two substantial improvements to our method. Firstly, the posterior distribution $p(a_{jk}, b_{jk} | \mathcal{D}_j, \mathcal{D}_k)$ now provides a distributional measure of the uncertainty in the shifting parameters. We may either plot this distribution to understand the certainty in the inferred shifts, or we may summarize it by computing the concavity of its logarithm around its maximum. Secondly, this formalism allows one to directly encode prior expectations regarding the shift parameters in order to break degeneracy in certain circumstances. For instance, some authors have noted that in the case of ambiguous shifting in time-cure superposition, it is sensible to minimize the extent of horizontal shifting, which may be encoded by a simple Gaussian prior in $\ln a_{jk}$ centered about zero \cite{larsen2008kaj}:
\begin{equation}
    - \ln p(a_{jk}, b_{jk}) \propto \lambda^2 (\ln a_{jk})^2. \label{eq:gaussian_prior}
\end{equation}
When employing the negative logarithm of the posterior distribution in estimation, this prior is equivalent to introducing $L^2$-regularization in the maximum likelihood framework \cite{Hoerl1970}. The parameter $\lambda$ in this case may be selected using one of many methods for hyperparameter optimization \cite{gelman2013,zhang1993}. In this work, we employ Monte Carlo cross-validation for optimization of $\lambda$ \cite{xu2004}, which will be discussed briefly in the next section.

The maximum likelihood framework developed previously is extended to maximum \textit{a posteriori} estimation simply by replacing the likelihood function within the objective in equation \ref{eq:max_likelihood_objective} with the posterior distribution defined in \ref{eq:posterior}. The minimizers $\hat{a}_{jk}$ and $\hat{b}_{jk}$ upon this substitution become maximum \textit{a posteriori} estimates, which are accompanied by posterior distributions that enable us to quantify their uncertainty. For these posterior distributions to be meaningful, they must be normalized such that their integral over their entire domain is unity.

When our prior expectations about the shift factors are limited, we may employ a uniform prior over $a_{jk}$ and $b_{jk}$. This uniform prior is the maximum entropy (or equivalently, minimum information) prior possible when $a_{jk}$ and $b_{jk}$ are bounded. With a uniform prior (i.e. assigning a constant value of $p(a_{jk}, b_{jk})$), the posterior distribution is equal to the likelihood function, and therefore the maximum likelihood and maximum \textit{a posteriori} estimates are equivalent. In the examples which follow, we assume a uniform prior unless otherwise noted, and bound the region of shift parameters to be that which results in nonzero overlap between data sets.

\subsection{Monte Carlo Cross-Validation}

In cases where a uniform prior over the shift parameters is not appropriate, we instead employ a prior, such as the Gaussian prior in equation \ref{eq:gaussian_prior}, that typically contains one or more hyperparameters ($\lambda$ in the case of equation \ref{eq:gaussian_prior}). This prior is meant to encode an expectation for the shift factors; for example, in cases where a wide range of shift factors result in equivalent superposition of two data sets, a Gaussian prior favors the minimum extent of shifting that also superposes the data. In order for the prior to work effectively, however, an appropriate value of the hyperparameter must be selected, without advanced knowledge of the `tightness' of the prior distribution that will properly balance the preference for minimal shifting against the quality of superposition. To find such a value, we employ one of many methods for hyperparameter optimization.

A general class of common methods for hyperparameter optimization is cross-validation, in which a fraction of the original data set is left out (the `validation' set), and the model is fit to the remaining data (the `training' set) \cite{gelman2013, zhang1993}. The optimal hyperparameter value is typically chosen as that value which, when used to fit the training data, results in the best predictive performance over the validation set. The methods for partitioning the data into validation and training sets vary. Here, we use a technique called Monte Carlo cross-validation (MCCV), in which training data are selected randomly and without replacement from the original data set, with a fraction $\alpha$ of the data retained for training and the remaining fraction $(1 - \alpha)$ held out for validation \cite{xu2004}. To apply MCCV to our method, we randomly partition each data set (i.e. the data at a particular state, $t_j$) individually, thus building an ensemble training set consisting of training data from all states, and a corresponding ensemble validation set. The sampling and validation are repeated for $K$ MCCV `folds' to obtain an estimate of the true cross-validation error associated with a particular hyperparameter value.

A single MCCV fold consists of partitioning the data, applying our automated superposition algorithm using a Gaussian prior with a particular value of $\lambda$ to the training data in order to obtain a master curve, fitting this master curve to its own Gaussian process model, and then evaluating the joint NLL of observing the validation data from this model, denoted $\mathcal{L}_k(\lambda)$. We repeat this for $K$ MCCV folds, and compute the average joint NLL over all folds, $\mathcal{L}(\lambda) = K^{-1}\sum_k \mathcal{L}_k(\lambda)$. In the limit that $K$ is large, the value of $\lambda$ that minimizes $\mathcal{L}(\lambda)$, denoted $\lambda_{m}$, is optimal. However, evaluating a large number of MCCV folds is expensive, thus the estimate of $\lambda_{min}$ is subject to some variation due to the sampling procedure. With a limited number of folds, a more conservative estimate of the optimal $\lambda$ is the maximum value that is within one standard error of the joint NLL loss at $\lambda_{min}$ \cite{hastie2009, silmore2019}:
\begin{equation}
    \scriptstyle \hat{\lambda} = \max \big\{\lambda \big\rvert \mathcal{L}(\lambda) \leq \mathcal{L}(\lambda_{m}) + K^{-1}\sqrt{\sum_k (\mathcal{L}_k(\lambda_{m}) - \mathcal{L}(\lambda_{m}))^2} \big\}.
\end{equation}
When hyperparameter optimization is complete, we then apply the superposition approach with a Gaussian prior and optimal hyperparameter $\hat{\lambda}$ to the entire data set.

\section{Detailed examples drawn from rheology}
\label{sec:examples}

In this section, we present a number of illustrative examples taken from the extensive literature on the application of the method of reduced variables in order to demonstrate the performance of our data superposition methodology. These examples are presented in increasing order of complexity - beginning with horizontal-only shifting for time-temperature superposition, continuing to vertical and horizontal shifting for time-cure superposition, and then concluding with two examples of introducing non-multiplicative shifting: one in shear rate-packing superposition of a liquid-liquid emulsion, and another in time--age-time superposition of a physically aging soft glassy material. In each case, we highlight the accuracy of our method by comparing the results of our algorithm to results obtained by other means, and emphasize how the trained models may be interpreted to gain insight into the physics underlying the observed data. We also provide timing benchmarks, recorded on a 2019 MacBook Pro (2.4 GHz 8-Core Intel Core i9 processor, 64 GB RAM), to demonstrate the computational efficiency of our approach. The software implementation of our algorithm, which has been used to obtain the results in this Section, is available for download at \url{https://github.com/krlennon/mastercurves}.

\subsection{Time-Temperature Superposition of a Polymer Melt}

\begin{figure*}
    \centering
    \includegraphics[width=\textwidth]{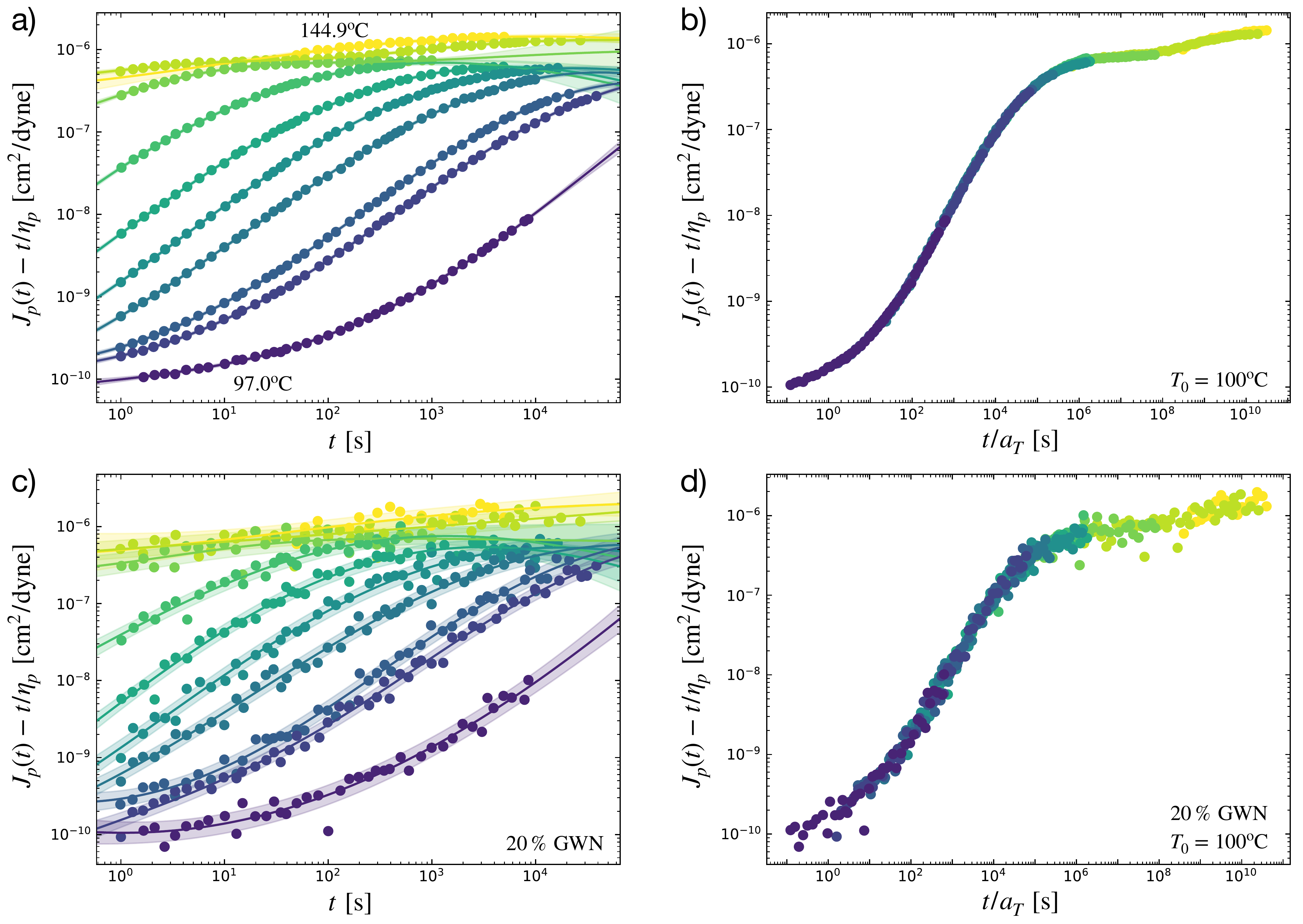}
    \caption{Automated superposition of recoverable creep compliance data acquired for a polystyrene melt at different temperatures. (a) The recoverable creep compliance data in laboratory time (circles), for temperatures of $T = 97.0^{\circ}\mathrm{C}$, $100.6^{\circ}\mathrm{C}$, $101.8^{\circ}\mathrm{C}$, $104.5^{\circ}\mathrm{C}$, $106.7^{\circ}\mathrm{C}$, $109.5^{\circ}\mathrm{C}$, $114.5^{\circ}\mathrm{C}$, $125.0^{\circ}\mathrm{C}$, $133.8^{\circ}\mathrm{C}$, and $144.9^{\circ}\mathrm{C}$, shown vertically in increasing order (digitized from \cite{plazek1965}). Solid lines and shaded regions show the posterior mean $m(t)$ and uncertainty bounds corresponding to one standard deviation, $m(t) \pm s(t)$, determined via Gaussian process regression. (b) Automatically constructed master curve using horizontal shifting, with a reference temperature of $T_0 = 100^{\circ}\mathrm{C}$. (c) The recoverable creep compliance with added 20\% relative Gaussian white noise, along with the associated posterior mean $m(t)$ and one standard deviation bounds $m(t) \pm s(t)$ determined by Gaussian process regression. (d) Automatically constructed master curve using horizontal shifting with 20\% relative Gaussian white noise added to the raw data.}
    \label{fig:tts}
\end{figure*}

Time-temperature superposition is an ubiquitous method for analyzing data pertaining to the thermomechanical response of polymeric materials \cite{ferry1980, markovitz1975}. Data at different temperatures may be used to construct master curves that span tens of orders of magnitude in time or frequency, a range far beyond that achievable in a single experiment. Here, we analyze canonical creep data taken for an entangled melt of polystyrene by Plazek \cite{plazek1965}. The digitized data, shown in Figure \ref{fig:tts}a, were rescaled to the recoverable creep compliance by the original author, who first subtracted out a response mode proportional to the measured background viscosity $\eta(T)$, then multiplied the result by a temperature and density dependent factor:
\begin{equation}
    J_p(t;T) - t/\eta_p(T) \equiv \left(\frac{T\rho}{T_0\rho_0}\right)\left[J(t;T) - t/\eta(T)\right],
\end{equation}
where $T$ and $\rho$ are the temperature and density describing a specific data set, and $T_0$ and $\rho_0$ are the reference temperature (373.2 K) and the density at the reference temperature. Because the vertical rescaling is accomplished using independently measured quantities such as $\rho(T)$, superposition of data sets at different temperatures requires only horizontal, multiplicative shifting.

In the notation developed in the previous section, the individual data sets are described by the state parameter $t = T$, and the measured data points are $(x_i, C_i)_j = (t_i, J_p(t_i; T_j) - t_i/\eta_p(T_j))$. We take the logarithm of this data for Gaussian process regression and inference, and perform a grid search over values of the horizontal shift parameter, $a_{jk}$, with shifting performed pairwise in order of decreasing temperature. Including the time required to fit the GP models, the superposition algorithm converges in 1.6 seconds. The master curve produced by the algorithm is shown in Figure \ref{fig:tts}b, with a reference temperature selected to be $T_0 = 100^{\circ}\mathrm{C}$ as in \cite{plazek1965}. Visually, the data have been brought into registry to create a single master curve with minimal deviations. Comparing Figure \ref{fig:tts}b with Figure 4 of \cite{plazek1965} confirms that the automatically and manually generated master curves are visually similar, and span nearly the same range in the shifted time coordinate.

A more quantitative comparison of the automated and manual shifting results is seen in Table \ref{tab:tts}, which lists the manually inferred shifts from \cite{plazek1965} and the maximum \textit{a posteriori} estimates from our automated method. The manual and automated shifts are quantitatively close, particularly at lower temperatures, where the slope of the data is greater. At higher temperatures, however, our automated procedure identifies that smaller shifts than those determined manually are required to optimally superpose the data. Moreover, we have obtained uncertainty estimates on these shift factors by computing the inverse Hessian of the joint negative log-posterior over each shift factor $a_T$, under the assumption that the posterior is approximately Gaussian. With this assumption, each diagonal element of the inverse Hessian matrix, $H_T^{-1}$, is related to the variance $\sigma_T^2$ in the inferred value of the corresponding $a_T$ (relative to the value of $a_T$ inferred at the next lower temperature, due to pairwise shifting): $H_T^{-1} = \sigma_T^2$. The shift factor at $T = 97.0^{\circ}\mathrm{C}$ is taken from \cite{plazek1965} and assumed to be exact, and the relative uncertainty in the shifts is accumulated at successively higher temperatures, thus the relative uncertainty in the shift factors increases monotonically with temperature. In Table \ref{tab:tts}, these uncertainty estimates are shown to be quite small relative to the inferred $a_T$. Thus the automatically generated results are both accurate, and precise.

Finally, we test the robustness of our approach to the presence of noise in the underlying data by adding synthetic Gaussian white noise to the raw data. The data with added 20\% Gaussian white noise are shown in Figure \ref{fig:tts}c, along with the posterior mean and one standard deviation uncertainty bounds of the regressed Gaussian process models. The Gaussian process models naturally incorporate the noise in the data via larger standard deviations, leaving the posterior mean function largely unaffected, as compared to the case with no added noise. Thus, the superposition of the noisy data, depicted in Figure \ref{fig:tts}d, is visually similar to that in the noise-free case, creating a master curve that spans the same range in the temporal coordinate, and maintains the same shape. The inferred shift factors for the added noise case are listed in Table \ref{tab:tts} as well, and remain close to both the noise-free and manually inferred values. Although these maximum \textit{a posteriori} estimates are not substantially affected by a relatively low signal-to-noise ratio, the increased uncertainty in the learned Gaussian process models broadens the posterior distributions over the horizontal shift factors, leading to increased uncertainty bounds. In particular, the uncertainties of the inferred shifts in the added noise case are between one and two orders of magnitude greater than those for the noise-free case. These results confirm the robustness of our automated shifting technique to noise in the underlying data, and demonstrate that this noise is appropriately transferred to uncertainty in the inferred shifts by virtue of our Bayesian approach.

\begin{table}[t!]
    \centering
    \begin{tabular}{c|c|c|c}
        \begin{tabular}{c} T \\ $(^{\circ}\mathrm{C})$ \end{tabular} & \begin{tabular}{c} $a_T$ \\ (Manual) \end{tabular} & \begin{tabular}{c} $a_T$ \\ (Automated) \end{tabular} & \begin{tabular}{c} $a_T$, 20\% GWN \\ (Automated) \end{tabular} \\\hline
        97.0 & $1.35 \cdot 10^1$ & $ 1.349 \cdot 10^{1}$ & $ 1.35 \cdot 10^{1}$ \\
        100.6 & $6.03 \cdot 10^{-1}$ & $(5.98 \pm 0.05) \cdot 10^{-1}$ & $(6.2 \pm 0.3) \cdot 10^{-1}$ \\
        101.8 & $2.60 \cdot 10^{-1}$ & $(2.84 \pm 0.03) \cdot 10^{-1}$ & $(3.1 \pm 0.2) \cdot 10^{-1}$ \\
        104.5 & $4.12 \cdot 10^{-2}$ & $(4.22 \pm 0.05) \cdot 10^{-2}$ & $(4.4 \pm 0.3) \cdot 10^{-2}$ \\
        106.7 & $1.10 \cdot 10^{-2}$ & $(1.14 \pm 0.02) \cdot 10^{-2}$ & $(1.20 \pm 0.09) \cdot 10^{-2}$ \\
        109.5 & $2.63 \cdot 10^{-3}$ & $(2.71 \pm 0.05) \cdot 10^{-3}$ & $(2.7 \pm 0.2) \cdot 10^{-3}$ \\
        114.5 & $3.24 \cdot 10^{-4}$ & $(3.3 \pm 0.1) \cdot 10^{-4}$ & $(3.3 \pm 0.6) \cdot 10^{-4}$ \\
        125.0 & $1.35 \cdot 10^{-5}$ & $(1.5 \pm 0.1) \cdot 10^{-5}$ & $(1.2 \pm 0.5) \cdot 10^{-5}$ \\
        133.8 & $2.14 \cdot 10^{-6}$ & $(1.4 \pm 0.1) \cdot 10^{-6}$ & $(1.1 \pm 0.6) \cdot 10^{-6}$ \\
        144.9 & $3.89 \cdot 10^{-7}$ & $(1.6 \pm 0.2) \cdot 10^{-7}$ & $(1.2 \pm 0.7) \cdot 10^{-7}$
    \end{tabular}
    \caption{Manually and automatically inferred horizontal shift factors for the recoverable creep compliance of an entangled polystyrene melt \cite{plazek1965}. Manual shift factors are those presented in \cite{plazek1965}. The automated shift factors represent the maximum \textit{a posteriori} estimates under a uniform prior, with the reported uncertainty representing one standard deviation, $\sigma$. All shifts are computed for a reference temperature of $100^{\circ}\mathrm{C}$ to match the convention in \cite{plazek1965}. The third column lists the automatically computed shift factors with no added noise, and the fourth column lists the automatically computed shift factors with 20 \% relative Gaussian white noise (GWN) added to the raw data.}
    \label{tab:tts}
\end{table}

This example represents a success of our method in interpretable machine learning. The learned master curve and shift factors together form a predictive model, which can be employed for subsequent material design or discovery. This model is not all that is learned from our algorithm, however. The shift factors in Table \ref{tab:tts} may also be used to infer physical features of this polystyrene melt. For instance, fitting the shift factors to the Williams, Landel, and Ferry equation reveals the free volume and coefficient of thermal expansion of the melt \cite{plazek1965}.

\subsection{Time-Cure Superposition During Gelation}

\begin{figure*}
    \centering
    \includegraphics[width=\textwidth]{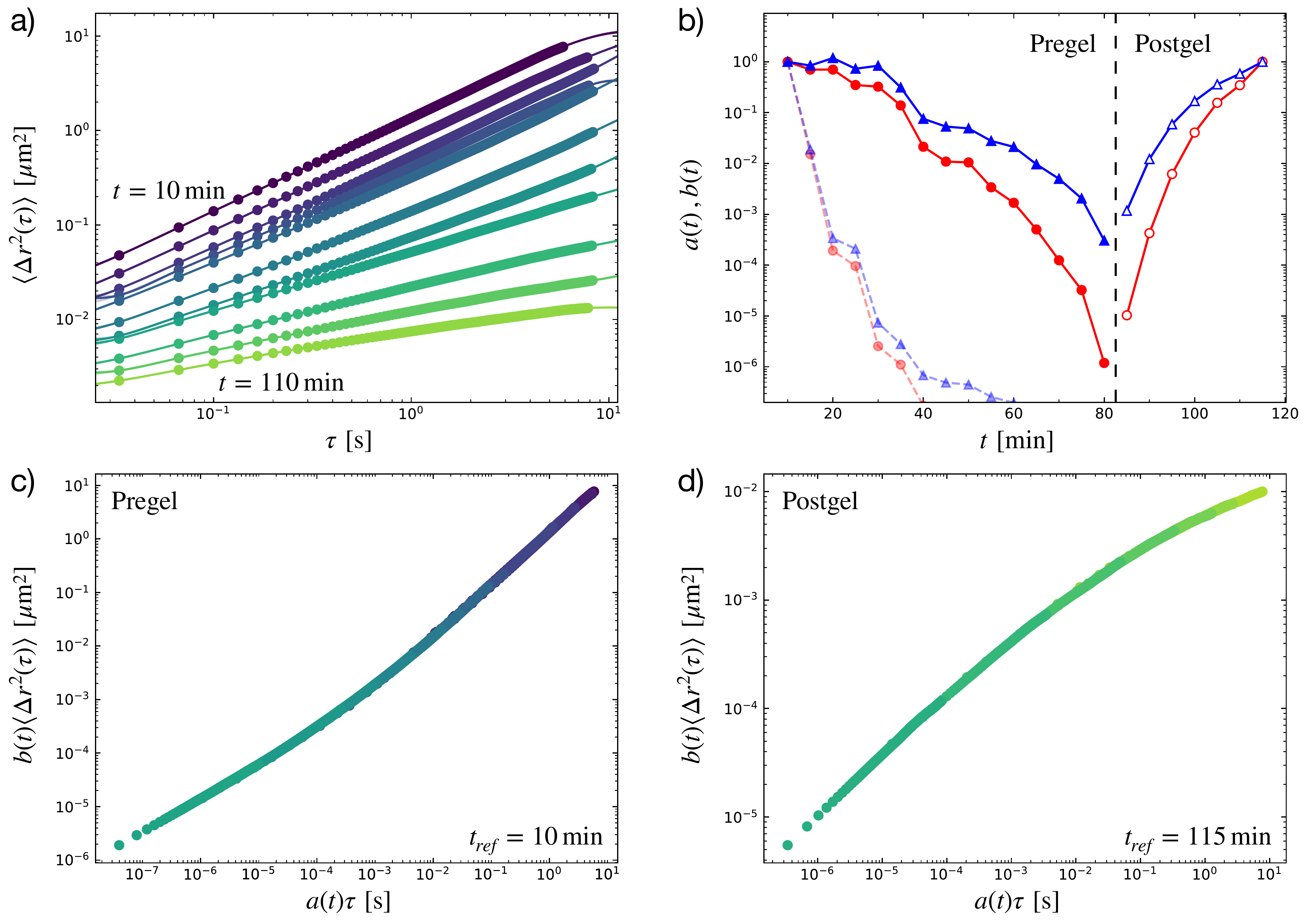}
    \caption{Automated superposition of mean-squared displacements measured by particle tracking passive microrheology in a peptide hydrogel undergoing physical gelation \cite{larsen2008}. (a) The mean squared displacement $\langle \Delta r^2 (\tau) \rangle$ as a function of lag-time $\tau$ (circles). Data was obtained for cure times $t \in [10, 115]$ minutes at 5 minute increments, though only data at 10 minute increments are shown here, presented in vertically descending order. Solid lines and shaded regions show the posterior mean $m(\tau)$ and uncertainty bounds corresponding to one standard deviation, $m(\tau) \pm s(\tau)$, determined via Gaussian process regression. (b) Maximum \textit{a posteriori} estimates of the horizontal (red circles) and vertical (blue triangles) shift factors. Lightly shaded points and lines represent values inferred under a uniform prior. Darker shaded points and lines represent values inferred with a Gaussian prior over $a(t)$ with hyperparameters optimized pairwise using Monte Carlo cross-validation. The shifts are optimally partitioned into a pregel curve and postgel curve. (c) Pregel master curve obtained with an optimized Gaussian prior over the shift factors $a(t)$. (d) Postgel master curve obtained with an optimized Gaussian prior over $a(t)$.}
    \label{fig:gelation}
\end{figure*}

The previous time-temperature superposition example demonstrated the effectiveness of our algorithm for multiplicative shifting of self-similar curve segments along the abscissa only. This is an important and widely applicable reference case; however, many instances of data superposition involve a rescaling of both the abscissa and the ordinate with changes in the state parameter. Simultaneous shifting along both axis complicates the shifting problem, in particular because it may result in a manifold of candidate shift parameters in two dimensions. Many previous methods for automatic generation of master curves cannot accommodate this more complicated case, either due to explicit limitations in their objective functions, or because they are not able to efficiently explore the two-dimensional domain of shift parameters. Our algorithm, however, extends naturally to simultaneous shifting on both axes, and its efficiency makes this two-dimensional optimization problem computationally tractable. Moreover, in the majority of circumstances involving rheological data, data are logarithmically scaled, so we may leverage the analytical solution for the optimal vertical shift presented in Section \ref{sec:log_shifting} to optimize shifts in two dimensions with minimal additional computational effort.

In this section, we present an example which encompasses simultaneous vertical and horizontal multiplicative shifting. The data set in this example represents a collection of mean-squared displacement versus lag-time trajectories that were obtained by particle tracking passive microrheology in a peptide hydrogel undergoing gelation \cite{larsen2008}. The state parameter in this data set represents the time $t$ since initial sample preparation (i.e., the `cure time'), ranging from 10-115 minutes in five minute increments. A subset of this data (those trajectories collected at 10 minute increments) is shown in Figure \ref{fig:gelation}a. Because the uncertainty characteristics of microrheological experiments differ from those of bulk rheology, we fit these data to Gaussian process models with $\sigma_u = 0$, to let the noise level be inferred solely from the data.

There are multiple particularly interesting features of this data set. One feature of importance, which has been noted in previous time-cure superposition studies \cite{Suman2021}, is that the shift factors tend to increase with distance from the gel point, $t_c$. This results in the creation of two master curves from the data set -- one for the pregel states $t < t_c$ and one for the postgel states $t > t_c$. The gel point can be inferred by determining the binary partition of the data leading to the optimal superposition into two master curves \cite{larsen2008}.

To construct the pregel and postgel master curves, and determine the optimal partition of the data set, we apply our superposition algorithm both in the forward (increasing $t$) and backward (decreasing $t$) directions, enforcing that the horizontal shifts decrease monotonically, and recording the pairwise posterior loss for each successive pair of curves in each direction. The total loss across a master curve is the sum of all pairwise losses; thus we may determine the optimal partition of the data to two master curves as that which minimizes the joint sum of pairwise losses associated with both curves. Specifically, if $\mathcal{L}_f(t_k)$ represents the loss incurred by adding the curve at state $t_k$ to the pregel (forward) master curve, and $\mathcal{L}_b(t_k)$ represents the loss incurred by adding that curve to the postgel (backward) master curve, then:
\begin{equation}
    t_c = \underset{t_c}{\arg\min} \sum_{t_k < t_c}\mathcal{L}_f(t_k) + \sum_{t_k > t_c}\mathcal{L}_b(t_k). \label{eq:gel_point}
\end{equation}
Upon applying the algorithm to this data set, however, one notices immediately that certain early cure time data sets are nearly linear trajectories in the log-log plots. Thus, there exists for these states a manifold of shifts resulting in nearly degenerate loss when applying a uniform prior to both the horizontal and vertical shift factors. In this case, the algorithm tends to prefer aligning curves at their ends, in order to minimize the overlap between the low-uncertainty (data-dense) regions of the underlying GP models, where deviations between curves is most heavily weighted in the joint NLL loss. This behavior results in extreme values of the shift factors $a(t)$ and $b(t)$. This case of excessive shifting is demonstrated by the horizontal and vertical shifts inferred with a uniform prior, shown as the lightly shaded points in Figure \ref{fig:gelation}b. For early times, these shifts decrease precipitously, even when no shifting at all would result in a similar likelihood of the master curve.

We may remedy these extreme shift factors by introducing a more informative prior over the shift factors, specifically the Gaussian prior presented in equation \ref{eq:gaussian_prior}. This prior is aligned with physical intuition, as previous authors have noted that in the case of ambiguous shifts it is sensible to limit shifts along the abscissa in favor of shifts along the ordinate, corresponding to an increased viscosity of a solution undergoing gelation \cite{larsen2008kaj}. For each successive pair of data sets, we optimize the hyperparameter $\lambda$ for this prior using MCCV. For this example, it was found that $\alpha = 0.1$, $K = 20$, and a grid search over 31 values of $\lambda \in [0.01, 10]$ (spaced logarithmically) were sufficient.

The darker shaded markers and lines in Figure \ref{fig:gelation}b represent the maximum \textit{a posteriori} estimates of the horizontal (red circles) and vertical (blue triangles) shift factors, $a(t)$ and $b(t)$ respectively, inferred from a Gaussian prior with optimized $\hat{\lambda}$. The introduction and optimization of this prior has clearly resulted in moderate values of shifts when superposing two nearly linear data sets (i.e. data obtained at early cure times), while still allowing for larger shifts when necessary (i.e. later cure times). The inferred shifts now closely resemble those inferred manually (c.f. Figure 3a in \cite{larsen2008}). Importantly, our algorithm and the condition specified in equation \ref{eq:gel_point} has also reproduced the manual determination of the gel point \cite{larsen2008}, where the optimal partition of the data set to two master curves occurs with $t_c$ between 80 and 85 minutes. This sensitive determination of a critical point demonstrates the capability of our automated shifting algorithm to detect meaningful, physically relevant features in rapidly evolving rheological data. The resulting pregel and postgel master curves are shown in Figure \ref{fig:gelation}c and d, both of which show excellent superposition, as well as close agreement with the manually constructed master curves reported in \cite{larsen2008}. Moreover, the shift factors depicted in Figure \ref{fig:gelation}b may be interpreted in terms of the divergence in the longest relaxation time and creep compliance of the hydrogel near the gel point. The exponents in this power-law divergence may be fit from the learned shift factors and compared to predicted results for certain universality classes of percolated gels, providing insight into the nature of hydrodynamic interactions in these evolving hydrogels \cite{larsen2008,Joshi2018}.

\subsection{Shear Rate-Volume Fraction Superposition of an Emulsion}

\begin{figure*}[t!]
    \centering
    \includegraphics[width=\textwidth]{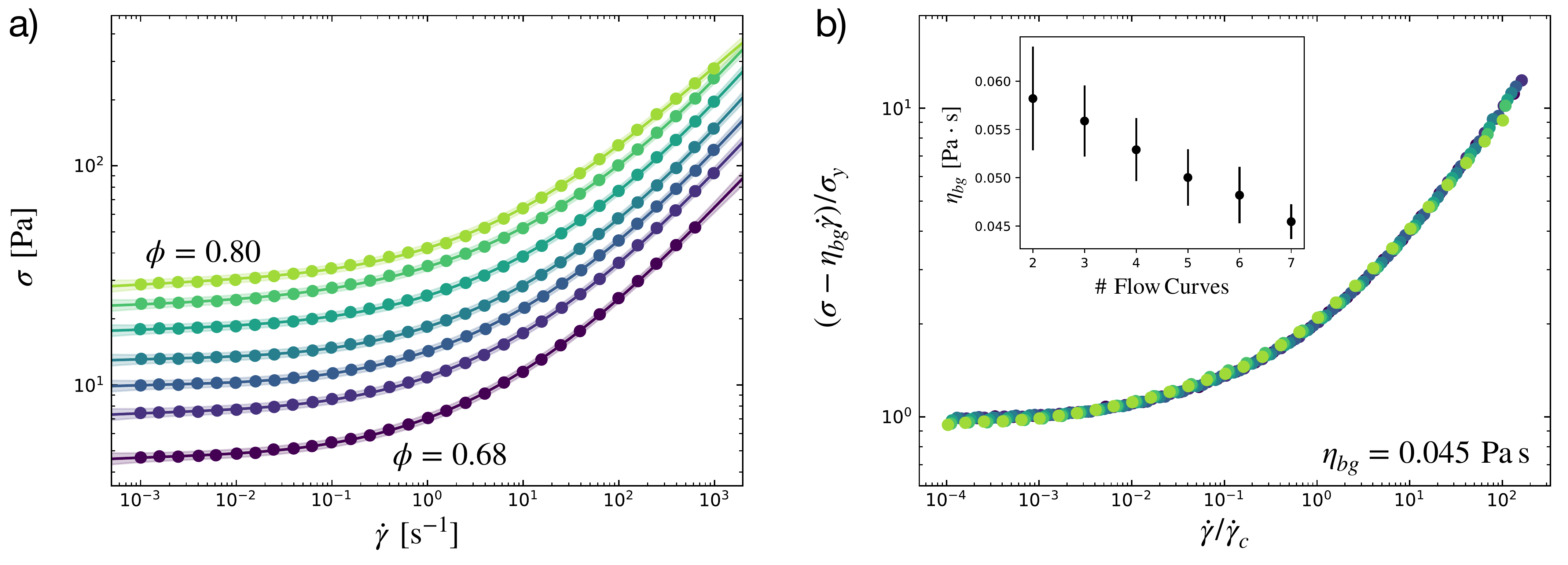}
    \caption{Automated superposition of steady flow data for castor oil-in-water suspensions with varying oil volume fraction. (a) The steady shear stress $\sigma$ measured over a range of steady shear rates $\dot{\gamma}$ (circles), for oil volume fractions of $\phi$ = 0.68, 0.70, 0.72, 0.74, 0.76, 0.78, and 0.80, shown vertically in increasing order (digitized from \cite{dekker2018}). Solid lines and shaded regions show the posterior mean $m(\dot{\gamma})$ and uncertainty bounds corresponding to one standard deviation, $m(\dot{\gamma}) \pm s(\dot{\gamma})$, determined via Gaussian process regression. (b) Automatically constructed master curve using subtraction of a state-independent purely viscous contribution to the stress with $\eta_{bg} = 4.5 \times 10^{-2}$ Pa$\cdot$s, followed by horizontal and vertical shifting. The reference state is taken as $\phi = 0.68$ with $\sigma_y = 4.7$ Pa and $\dot{\gamma}_c = 4.7$ s$^{-1}$. Inset shows the optimal value and estimated uncertainty of $\eta_{bg}$ inferred by applying the method to an increasing number of flow curves from (a), averaged over all possible combinations.}
    \label{fig:emulsion}
\end{figure*}

In Section \ref{sec:non_multiplicative}, we discussed how our automated, maximum \textit{a posteriori} shifting approach can be extended beyond simple multiplicative vertical and horizontal shifting. Due to the efficiency of the optimization over multiplicative shifts, for example, it is straightforward and still computationally tractable to perform other non-multiplicative coordinate transformations in an outer optimization problem. Moreover, this optimization hierarchy fits naturally within a Bayesian framework, so that the posterior distribution can easily incorporate a new parameter from the outer optimization problem. In this section,  we demonstrate one such problem: the superposition of steady flow curves obtained for castor oil-in-water emulsions at varying oil volume fractions \cite{dekker2018}.

Recently, a three-component model has been proposed to describe these steady flow curves, composed of elastic, plastic, and viscous contributions to the total shear stress in the emulsion \cite{caggioni2020}:
\begin{equation}
    \sigma(\dot{\gamma};\phi) = \sigma_y(\phi) + \sigma_y(\phi)\left(\frac{\dot{\gamma}}{\dot{\gamma}_c(\phi)}\right)^{1/2} + \eta_{bg}\dot{\gamma},
\end{equation}
where the yield stress $\sigma_y(\phi)$ and critical shear rate $\dot{\gamma}_c(\phi)$ both are assumed to vary with the oil volume fraction $\phi$, while the background viscosity $\eta_{bg}$ is typically assumed to be independent of $\phi$. This model assumes a true yield stress $\sigma_y$ and square root dependence of the plastic contribution on the shear rate. More generally, we may relax these assumptions to write the relationship:
\begin{equation}
    \sigma(\dot{\gamma};\phi) = \sigma_y(\phi)g\left(\frac{\dot{\gamma}}{\dot{\gamma}_c(\phi)}\right) + \eta_{bg}\dot{\gamma},
\end{equation}
where $g(\dot{\gamma}_r)$, with the reduced shear rate $\dot{\gamma}_r \equiv \dot{\gamma}/\dot{\gamma}_c(\phi)$, captures the nonlinear viscoplastic response of the emulsion. For example, an ideal Herschel–Bulkley fluid would be described by $\eta_{bg} = 0$ and $g(\dot{\gamma}_r) = 1 + \dot{\gamma}_r^n$. The parametric self similarity of this model is now clear, as the above equation is of the form of equation \ref{eq:master_curve}. To obtain the master curve $g(\dot{\gamma}_r)$, we must first subtract any purely viscous mode $\eta_{bg}\dot{\gamma}$ (e.g. arising from the background solvent) from $\sigma(\dot{\gamma};\phi)$, then apply vertical and horizontal multiplicative shifting.

In Figure \ref{fig:emulsion}a, we show steady flow curves for castor oil-in-water emulsions at seven different oil volume fractions spaced evenly between 0.68 and 0.80, digitized from \cite{dekker2018} and each fit to a Gaussian process model. Optimization of the volume fraction-dependent vertical and horizontal shift factors, $\sigma_y$ and $\dot{\gamma}_c$, as well as the volume fraction-independent background viscosity $\eta_{bg}$ proceeds hierarchically as described previously. In particular, we perform a grid search over 100 linearly spaced values in $\eta_{bg} \in [0, 0.1]$. At each value, we compute the optimal $\sigma_y$ and $\dot{\gamma}_c$ (selecting a reference state of $\phi = 0.68$ with $\sigma_y = 4.7$ Pa and $\dot{\gamma}_c = 4.7$ s$^{-1}$, as computed in \cite{caggioni2020}) using our algorithm for multiplicative-only shifting, and we record the total negative log posterior loss. Under a uniform prior, the maximum \textit{a posteriori} estimate of $\eta_{bg}$ corresponds to the numerical value minimizing this loss, and the uncertainty in $\eta_{bg}$ can be estimated from a finite-difference approximation for the Hessian of the recorded loss about the minimum.

\begin{figure}
    \centering
    \includegraphics[width=\columnwidth]{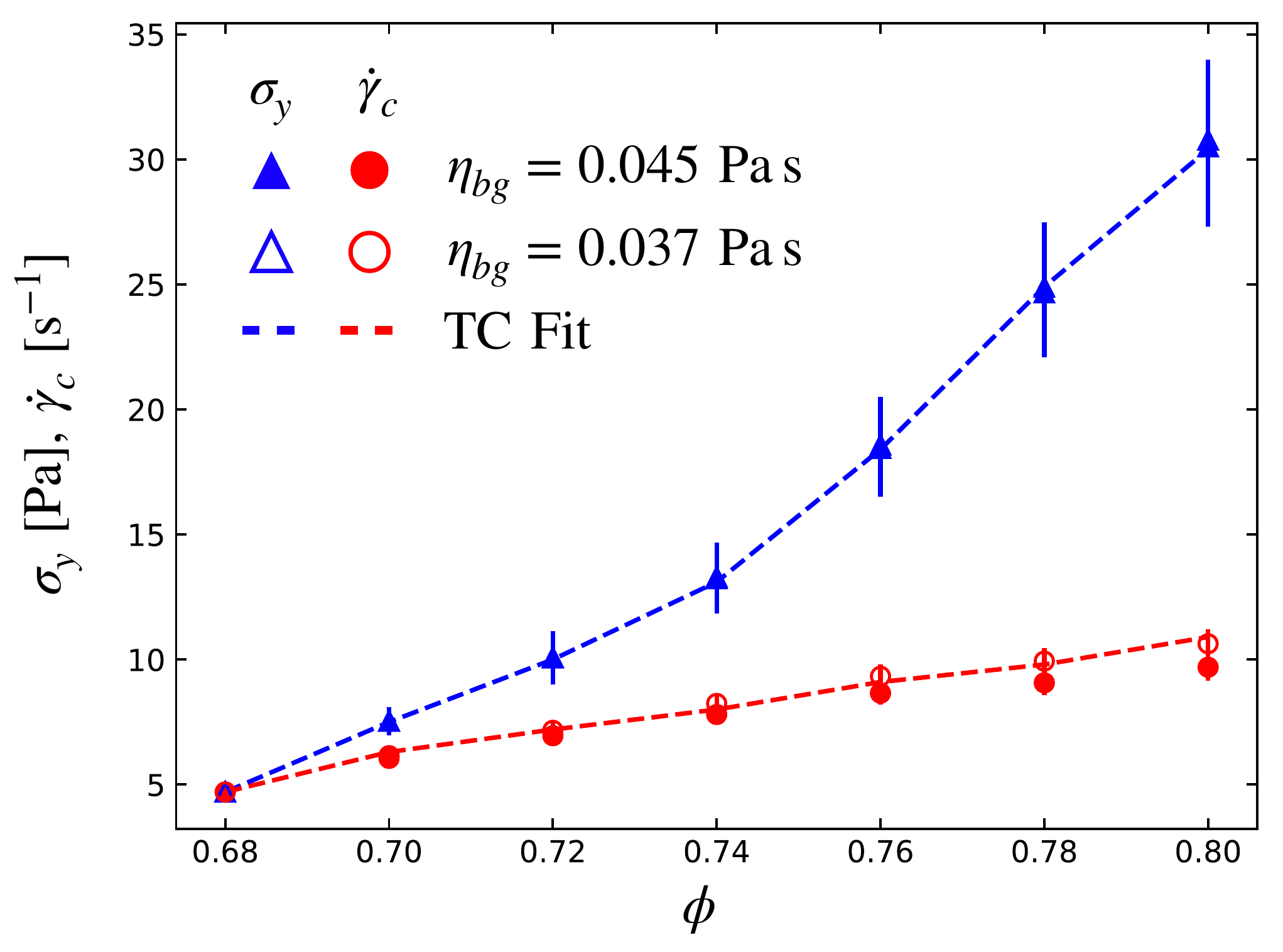}
    \caption{The critical shear rate and yield stress values for castor oil-in-water emulsions inferred by the automated algorithm with the maximum \textit{a posteriori} estimate of $\eta_{bg} = 4.5 \times 10^{-2}$ Pa$\cdot$s (filled symbols), and with $\eta_{bg} = 3.7 \times 10^{-2}$ Pa$\cdot$s (unfilled symbols), as well as values fit via the three-component (TC) model (dashed lines) \cite{caggioni2020}. Vertical bars depict one standard error in the estimates.}
    \label{fig:emulsion_shifts}
\end{figure}

Upon performing this optimization, we obtain a master curve with optimal vertical and horizontal shifts, as well as the maximum \textit{a posteriori} estimate of the background viscosity. At each value of $\eta_{bg}$ in the grid search, the optimization takes approximately 1.3 seconds to converge. The resulting master curve is presented in Figure \ref{fig:emulsion}b, which corresponds to a maximum \textit{a posteriori} estimate of $\eta_{bg} = (4.5 \pm 0.2) \times 10^{-2}$ Pa$\cdot$s. This estimate is only slightly greater than that obtained via direct fits of the three-component model, $\eta_{bg} = 3.7 \times 10^{-2}$ Pa$\cdot$s \cite{caggioni2020}, but now incorporates all data sets in its determination and is accompanied by an uncertainty estimate. To emphasize the benefit of employing the entire data set in robust estimation of $\eta_{bg}$, we perform the same optimization using pruned data sets, consisting of a subset of the seven flow curves in Figure \ref{fig:emulsion}a. The inset in Figure \ref{fig:emulsion}b presents the inferred value of $\eta_{bg}$, and its associated uncertainty, averaged over all possible combinations of these subsets of a given size. The value of $\eta_{bg}$ inferred with fewer flow curves deviates progressively from the value that is optimal for the entire data set, and has substantially higher uncertainty. That the parameter estimate becomes more precise when including more data highlights a salient feature of our methodology -- namely, that its inferences and predictions may be continually refined by including more data.

\begin{figure*}
    \centering
    \includegraphics[width=\textwidth]{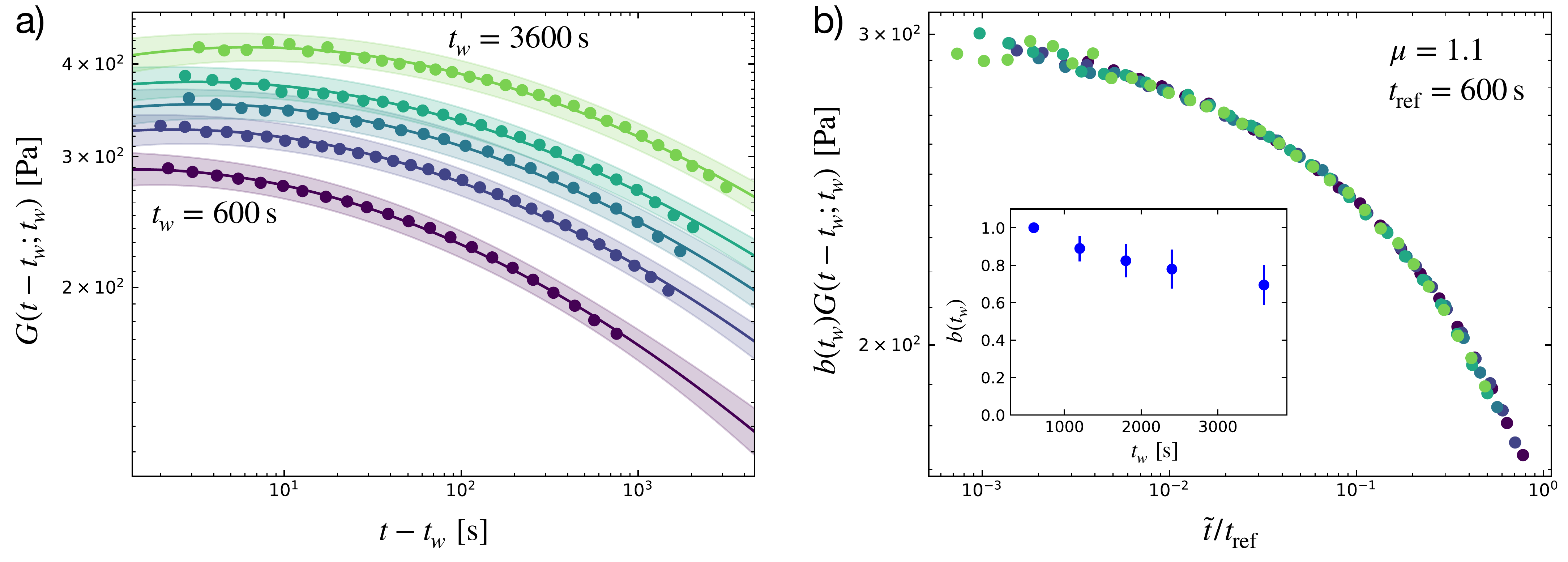}
    \caption{Automated superposition of stress relaxation data for a physically aging suspension of Laponite\textsuperscript{\textregistered}-RD clay with varying wait time $t_w$ between mixing and preshearing, and imposition of a step strain with $\gamma_0 = 0.03$ (which is always in the linear viscoelastic range). (a) The relaxation modulus $G(t-t_w;t_w)$ as a function of the time since the step strain, $t - t_w$, parametric in the wait time $t_w$. Data (circles) are shown for $t_w$ = 600 s, 1200 s, 1800 s, 2400 s, and 3600 s, shown vertically in increasing order (digitized from \cite{gupta2012}). Solid lines and shaded regions show the posterior mean $m(t - t_w)$ and uncertainty bounds corresponding to one standard deviation, $m(t - t_w) \pm s(t - t_w)$, determined via Gaussian process regression. (b) Automatically constructed master curve using transformation to the effective or material time interval $\tilde{t}/t_{\mathrm{ref}} = \left[\left(t/t_{\mathrm{ref}}\right)^{1-\mu} - \left(t_w/t_{\mathrm{ref}}\right)^{1-\mu}\right]/(1 - \mu)$ with $\mu = 1.1$, and subsequent vertical multiplicative shifting by a factor $b(t_w)$, with a reference wait time of $t_{\mathrm{ref}}$ = 600 s. The vertical shift factors $b(t_w)$ are shown in the inset, with vertical bars representing one-standard-error uncertainty estimates.}
    \label{fig:laponite}
\end{figure*}

The inferred horizontal and vertical shifts, corresponding to the critical shear rate $\dot{\gamma}_c$ and yield stress $\sigma_y$, are presented in Figure \ref{fig:emulsion_shifts}. The values determined via individual fits to the three-component model with $\eta_{bg} = 3.7 \times 10^{-2}$ Pa$\cdot$s are shown by dashed lines for clarity \cite{caggioni2020}. In parallel with the values determined by our automated shifting procedure for the maximum \textit{a posteriori} estimate of $\eta_{bg} = 4.5 \times 10^{-2}$ Pa$\cdot$s, we also present using hollow symbols the corresponding shifts inferred by our algorithm with a fixed value of $\eta_{bg} = 3.7 \times 10^{-2}$ Pa$\cdot$s to enable a direct comparison to the previously reported three-component values in \cite{caggioni2020}. The yield stresses (triangles) determined by each method are remarkably similar and unaffected by variation in $\eta_{bg}$. The critical shear rates are also quite similar for all cases. The maximum \textit{a posteriori} estimates for $\dot{\gamma}_c(\phi)$ with $\eta_{bg} = 3.7 \times 10^{-2}$ Pa$\cdot$s are very close to those from the three-component fit -- within one standard error for all $\phi$ -- confirming that our method produces quantitatively similar results to parameter estimation within a constitutive model. The optimal values of $\dot{\gamma}_c(\phi)$ with $\eta_{bg} = 4.5 \times 10^{-2}$ Pa$\cdot$s are slightly lower, demonstrating some sensitivity of the method to the choice of background viscosity and again highlighting the importance of including the entire data set for parameter estimation.

Within our interpretable machine learning framework, the learned shift factors in Figure \ref{fig:emulsion_shifts} may be explained in terms of a power-law dependence of the yield stress and critical shear rate on the distance to jamming \cite{dekker2018}. The exponents in such a dependence typically suggest that these emulsions belong to a certain universality class describing a transition from liquid-like to solid-like dynamics \cite{Paredes2013}.

\subsection{Time--Age-Time Superposition of an Aging Soft Glassy Material}

The previous examples encompass multiple cases of vertical and horizontal multiplicative shifting, with the additional feature of state-independent non-multiplicative shifting in the last example. These linear coordinate transformations are ubiquitous in the analysis of rheological data; however, there are other more complex material responses that require other, nonlinear coordinate transformations. One example is that of power-law rheological aging in soft glassy materials, in which the modulus and relaxation dynamics become parametrically self-similar only in an effective time domain \cite{gupta2012,struik1977}:
\begin{equation}
    \frac{\tilde{t}}{t_{\mathrm{ref}}} = \frac{1}{1 - \mu}\left[\left(\frac{t}{t_{\mathrm{ref}}}\right)^{1-\mu} - \left(\frac{t_w}{t_{\mathrm{ref}}}\right)^{1-\mu}\right]. \label{eq:power_age}
\end{equation}
The effective time interval $\tilde{t}$ is related by a power law in a state-independent parameter $\mu$ (provided that $\mu \neq 1$) to the laboratory time $t$ and the `wait time' $t_w$ (which represents roughly the elapsed time between material preparation and the beginning of a measurement), with $t_{\mathrm{ref}}$ representing the selected reference state.

For a soft glassy material undergoing power law aging, one can measure the relaxation modulus $G(t - t_w; t_w)$ by applying step strains at various times $t_w$. In this case, the state parameter $y$ is the wait time $t_w$. However, due to the power-law dilation of time described by equation \ref{eq:power_age}, the system is not time-translation-invariant \cite{Joshi2018,Fielding2000}, and therefore the relaxation modulus is no longer related to a master curve by the relation specified in \ref{eq:master_curve} with $x = t$. Instead, it is self-similar in the effective time domain:
\begin{equation}
    g(\tilde{t}) = b(t_w)G(t - t_w; t_w),
\end{equation}
where $\tilde{t}$ depends on $t$, $t_w$, and the exponent $\mu$ as per equation \ref{eq:power_age}. Given this similarity relation, we may apply the same hierarchical optimization approach as in the previous section to the superposition of stress relaxation data in an aging soft material. In an outer optimization loop, we perform a grid search over $\mu$, transforming the data into the effective time domain $\tilde{t}(\mu)$ for each value of $\mu$ in the search. In the inner loop, the optimal vertical shifts are computed analytically using the approach described in Section \ref{sec:log_shifting}.

In Figure \ref{fig:laponite}a we show stress relaxation data taken on an aqueous suspension of Laponite\textsuperscript{\textregistered}-RD clay at 2.8 wt.\%, digitized from \cite{gupta2012}. This clay suspension is known to exhibit soft glassy characteristics, including power-law aging. Data are obtained for five different wait times $t_w$ between 600 s and 3600 s. We fit each data set to a Gaussian process model, and perform a grid search over 100 linearly spaced values of the state-independent parameter $\mu \in [0.1, 1.9]$, representing a range of values typical for aging clay suspensions \cite{Joshi2018}. At each value of $\mu$ in the grid search, optimization of the multiplicative vertical shift factors proceeds analytically and in a pairwise fashion. Finally, starting from the optimal value of $\mu$ determined by the grid search, we perform gradient descent to refine the estimation, and compute the Hessian for uncertainty analysis. In total, the grid search and subsequent gradient descent converges in just under one second, producing the maximum \textit{a posteriori} estimate of $\mu = 1.1 \pm 0.1$. The resulting master curve for a chosen reference time of $t_{\mathrm{ref}} = 600$ s is presented in Figure \ref{fig:laponite}b.

The inferred value of $\mu$ is virtually identical to that obtained manually -- $\mu \approx 1.1$ \cite{gupta2012}, but its determination comes with the additional benefits of an uncertainty estimate and analytic determination of $b(t_w)$ (shown in the inset of Figure \ref{fig:laponite}b). This demonstrates that our automated shifting algorithm effectively handles more complicated coordinate transformations, such as time dilation via power law aging, in a manner consistent with expert manual analysis. The precise determination of $\mu$ is also significant, because $\mu$ is often used to delineate ``hyperaging'' ($\mu > 1$) and ``sub-aging'' ($\mu < 1$) materials. Therefore, the estimate of $\mu = 1.1$ also serves to identify that this clay suspension is in the universality class of hyperaging materials, and this automated classification is consistent with the expert manual classification. By collating a series of such \emph{time--age-time superposition} (\textit{taTs}) measurements at different temperatures, the inferred value of $\mu$ and its dependence on temperature may be interpreted in terms of the microscopic yielding dynamics of soft glassy materials, and subsequently related to characteristic length scales and yielding energy barriers associated with the aging colloidal gel \cite{gupta2012}.

\section{Forward Predictions using Automated Shifting}
\label{sec:predictions}

In the previous section, we provided four canonical examples of automated construction of rheological master curves using real experimental data taken from the literature, demonstrating the robustness of our automated method to different types of data, types of shifting, and levels of noise. The automated shifting algorithm is thus a useful tool in the analysis of linear and nonlinear thermorheological data, as the inferred shifting parameters provide physical insight to key material properties -- such as the static yield stress of an emulsion, the gel point of a polymeric solution, or the rate of aging of a soft glassy material. Another potent application of this algorithm is to make reliable forward predictions of material rheology at a previously unstudied state. In this section, we present an example of one such forward prediction, using the castor oil-in-water flow curve data \cite{dekker2018}.

Within the mathematical framework employed in this work, the forward prediction problem is as follows. Given data for a material property $C(x;t)$ at a set of states $\{t_j\}$, including a reference state $t_k$, we infer optimal shifting parameters -- $\{a_{jk}\}$, $\{b_{jk}\}$, and $c$ in the scenario described by equation \ref{eq:master_curve} -- and apply these shifts to the data to construct a master curve approximating $g(z)$ in the reduced variable $z = a(t)x$. To make a forward prediction at a new state $t_l$, we must estimate the shift parameters $a_{lk}$ and $b_{lk}$ for that state, and then use equation \ref{eq:master_curve} to transform the inferred master curve $g(z)$ into a prediction of $C(x;t_l)$ at state $t_l$. Making predictions continuously in the independent variable $x$ requires a continuous approximation for $g(z)$, which may be obtained by fitting the automatically constructed master curve to a new Gaussian process model. Similarly, predicting $a_{lk}$ and $b_{lk}$ requires an interpolant between the inferred set of $\{a_{jk}\}$ and $\{b_{jk}\}$; this interpolant can also be treated using a Gaussian process model.

\begin{figure}
    \centering
    \includegraphics[width=\columnwidth]{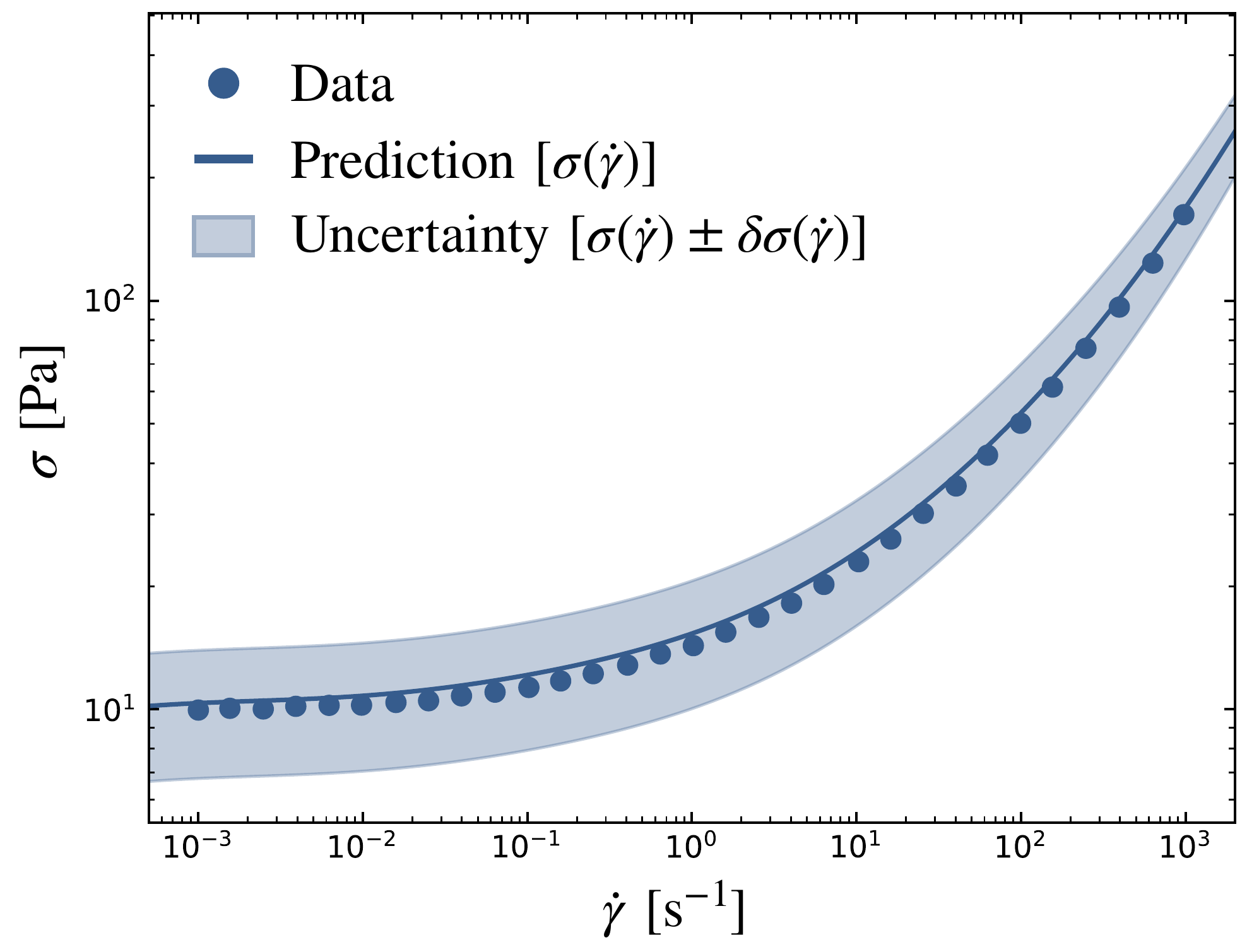}
    \caption{Forward predictions of the steady flow curve for a castor oil-in-water emulsion with oil volume fraction $\phi = 0.72$. The value of $\eta_{bg} = (4.5 \pm 0.2) \times 10^{-2}$ Pa$\cdot$s is inferred during construction of a master curve from the remaining data sets ($\phi$ = 0.68, 0.70, 0.74, 0.76, 0.78, 0.80). The yield stress $\sigma_y = 9 \pm 1$ Pa and critical shear rate $\dot{\gamma}_c = 6.7 \pm 0.1$ s$^{-1}$ are estimated from Gaussian process models fit to the automatically inferred shift factors at the remaining states. The inferred master curve is fit to a Gaussian process model, whose predictions are shifted to the $\phi = 0.72$ state using the predicted values of $\sigma_y$, $\dot{\gamma}_c$, and $\eta_{bg}$. The mean predicted values are shown with a solid line, and single standard deviation uncertainty bounds are shown with a shaded region. Experimental data for $\phi = 0.72$ from \cite{dekker2018} are shown by the solid circles.}
    \label{fig:prediction}
\end{figure}

To demonstrate this forward prediction procedure, we automatically construct a master curve for the castor oil-in-water data set shown in Figure \ref{fig:emulsion}, but excluding the data set at $\phi = 0.72$. The master curve constructed while deliberately omitting this data set out is still very similar to that depicted in Figure \ref{fig:emulsion}b, as are the inferred shift factors $\sigma_y$ and $\dot{\gamma}_c$ for the remaining volume fractions, and the estimate of $\eta_{bg} = (4.5 \pm 0.2) \times 10^{-2}$ Pa$\cdot$s. These shift factors are fit to Gaussian process models as a function of $\phi$, and these GPs are used to make a forward prediction of the yield stress and critical shear rate at $\phi = 0.72$. To ensure that the GPs fit to these shifts incorporate the uncertainty in the underlying shift factors, we add a white noise kernel to the GPs, with a fixed noise level for each individual volume fraction equal to the greatest variance of any shift factor (which are shown by the error bars in Figure \ref{fig:emulsion_shifts}). The predictions from the resulting GPs are: $\sigma_y = 9 \pm 1$ Pa and $\dot{\gamma}_c = 6.7 \pm 0.1$ s$^{-1}$, both in close agreement with the values obtained by automated shifting of the data set at $\phi = 0.72$ and the values determined from fits to the three-component model (c.f. Figure \ref{fig:emulsion_shifts}).

Finally, the master curve obtained from all data sets besides that for $\phi = 0.72$ is fit to its own GP model. The self-similar posterior mean function $g(\dot{\gamma}_r)$ and variance $\delta g(\dot{\gamma}_r)$ of this GP model are then computed for a range of reduced shear rates $\dot{\gamma}_r \equiv \dot{\gamma}/\dot{\gamma}_c$, and the self-similar results are shifted back to the original ($\sigma$, $\dot{\gamma}$) parameter space by first applying vertical and horizontal shifts with the estimated $\sigma_y$ and $\dot{\gamma}_c$ at $\phi = 0.72$, then adding in the viscous mode with the inferred optimal value of $\eta_{bg}$:
\begin{equation}
    \sigma(\dot{\gamma}) = \sigma_y g(\dot{\gamma}/\dot{\gamma}_c) + \eta_{bg}\dot{\gamma}.
\end{equation}
Uncertainty in the value of the vertical shift $\delta \sigma_y$, the horizontal shift $\delta \dot{\gamma}_c$, and the background viscosity $\delta \eta_{bg}$ can also be propagated to the total uncertainty expected in the prediction of the resulting flow curve:
\begin{align}
    (\delta \sigma)^2 =\,& g(\dot{\gamma}/\dot{\gamma}_c)^2 (\delta \sigma_y)^2 + \sigma_y^2 (\delta g(\dot{\gamma}/\dot{\gamma}_c))^2 \nonumber \\
    & + \left(\frac{\sigma_y g'(\dot{\gamma}/\dot{\gamma}_c) \dot{\gamma}}{\dot{\gamma}_c^2}\right)^2 (\delta\dot{\gamma}_c)^2 + \dot{\gamma}^2 (\delta \eta_{bg})^2,
\end{align}
with $g'(\dot{\gamma}_r) = dg(\dot{\gamma}_r)/d\dot{\gamma}_r$. Figure \ref{fig:prediction} depicts the forward prediction for the steady shear stress of an emulsion with $\phi = 0.72$. The expectation for the flow curve $\sigma(\dot{\gamma})$ is shown as a solid line, along with uncertainty bounds $\sigma \pm d\sigma$ in the shaded region, over six orders of magnitude in the imposed shear rate $\dot{\gamma}$. The predictions are compared to the original withheld data, shown with filled circles. The data fall nearly exactly on the mean predictions over the entire range of $\dot{\gamma}$, and in all cases are well within the uncertainty bounds, which represent a conservative estimate of the total uncertainty due to the built-in noise level of $\sigma_u = 0.04$ in the Gaussian processes. In this case, the automatically generated master curve, in combination with Gaussian process models for the shift factors as a function of $\phi$, serves as a highly effective predictive tool. Notably, these accurate predictions have been made without assuming or specifying a specific constitutive model for the master curve, $g(\dot{\gamma}_r)$. Rather, the predictions are \emph{data-driven}, using Gaussian process regression as a machine learning tool to represent the underlying master curve. Moreover, the predictions are statistical in nature; the uncertainty bounds in Figure \ref{fig:prediction} are similarly derived directly from the available data, and are obtained with minimal extra work. Viewed holistically, this example of forward prediction highlights many of the salient features of our data superposition approach, and demonstrates how it may be straightforwardly adapted as a predictive tool for material design and formulation considerations.

\section{Conclusions}

In this work, we have developed a data-driven, statistical method for the inference of material master curves from parametrically self-similar data sets. Such responses are observed ubiquitously across the field of soft materials science and have led to the development of effective but labor-intensive approaches that are collected under the generic term \emph{the method of reduced variables}. Our new approach is inherently flexible, as the negative log-posterior objective is independent of the transformations used to bring different data sets into registry and can therefore accommodate a wide variety of data-shifting and superposition transformations. It is also computationally efficient, with optimal values of the vertical multiplicative shifts of (logarithmically scaled) data computed analytically. This allows for quick one-dimensional horizontal shifting and even outer-loop optimization of state-independent parameters (such as we illustrated using the background viscosity in a model for emulsion rheology). The method is robust to the presence of noise, as noise is handled explicitly by the Gaussian process model to which the data is regressed. Finally, the uncertainty estimates of the shift parameters, together with continuous uncertainty bounds for forward predictions, are obtained automatically from the method at minimal added cost, due to the formulation of the superposition problem as one of statistical inference.

Our method for inferring master curves not only facilitates the automation of a popular method for rheological data analysis and extrapolation that has endured for more than 80 years, but also represents the development of a novel probabilistic, data-driven predictive and analytic tool. We may therefore readily extend this method to new material systems, refine its predictions with more extensive data sets, and, potentially, build an open shared library of predictive models that are unbiased by user preferences and preconceptions. Moreover, this automation has been accomplished without sacrificing the physical interpretability of the resulting models. As we have demonstrated, the learned models themselves often provide valuable insight to the underlying physics governing the material systems under study. In all fields of the physical sciences, the adoption of robust, open, and unbiased data-driven tools such as this will be critical in developing both accessible and reproducible scientific insight across large data sets.

\section*{Data Availability Statement}

The data used in this work, along with a software implementation of the proposed methodology and the code used in all demonstrations, are available on GitHub at \url{https://github.com/krlennon/mastercurves}.

\section*{Author Contributions}

Conceptualization: K.R.L., G.H.M., J.W.S.; Methodology: K.R.L, J.W.S.; Data Curation: K.R.L., G.H.M., J.W.S.; Visualization: K.R.L., G.H.M., J.W.S.; Writing – original draft: K.R.L., J.W.S.; Writing – review \& editing: K.R.L., G.H.M., J.W.S.; Software: K.R.L.

\section*{Funding Statement}

K.R.L. was supported by the U.S. Department of Energy (DOE) Computational Science Graduate Fellowship program under Grant No. DE-SC0020347.

\section*{Competing Interests}

The authors declare that no competing interests exist.

\bibliographystyle{ieeetr}
\bibliography{biblio.bib}

\begin{thebibliography}{10}

\bibitem{Carleo2019}
G.~Carleo, I.~Cirac, K.~Cranmer, L.~Daudet, M.~Schuld, N.~Tishby,
  L.~Vogt-Maranto, and L.~Zdeborov\'a, ``Machine learning and the physical
  sciences,'' {\em Rev. Mod. Phys.}, vol.~91, p.~045002, Dec 2019.

\bibitem{Butler2018}
K.~T. Butler, D.~W. Davies, H.~Cartwright, O.~Isayev, and A.~Walsh, ``{Machine
  learning for molecular and materials science},'' {\em Nature}, vol.~559,
  no.~7715, pp.~547--555, 2018.

\bibitem{Ferguson2017}
A.~L. Ferguson, ``{Machine learning and data science in soft materials
  engineering},'' {\em Journal of Physics: Condensed Matter}, vol.~30, no.~4,
  p.~43002, 2017.

\bibitem{Barenblatt2003}
G.~I. Barenblatt, {\em {Scaling}}.
\newblock Cambridge: Cambridge University Press, 2003.

\bibitem{leaderman1944}
H.~Leaderman, {\em {Elastic and Creep Properties of Filamentous Materials and
  Other High Polymers}}.
\newblock Washington, D.C.: The Textile Foundation, 1944.

\bibitem{tobolsky1945}
A.~V. Tobolsky and R.~D. Andrews, ``Systems manifesting superposed elastic and
  viscous behavior,'' {\em The Journal of Chemical Physics}, vol.~13, no.~1,
  pp.~3--27, 1945.

\bibitem{ferry1980}
J.~D. Ferry, {\em Viscoelastic Properties of Polymers}.
\newblock John Wiley \& Sons, 1980.

\bibitem{wagner1915}
K.~W. Wagner, ``{Dielektrische Eigenschaften von verschiedenen
  Isolierstoffen},'' {\em Elektrotechnische Zeitschrift}, vol.~12, pp.~120,123,
  135--137, 163,165, 1915.

\bibitem{lillo2003}
F.~Lillo, J.~D. Farmer, and R.~N. Mantegna, ``{Master curve for price-impact
  function},'' {\em Nature}, vol.~421, no.~6919, pp.~129--130, 2003.

\bibitem{deGennes1979}
P.~de~Gennes, {\em Scaling Concepts in Polymer Physics}.
\newblock Cornell University Press, 1979.

\bibitem{Wagner2016}
N.~Wagner and J.~M. Rondinelli, ``Theory-guided machine learning in materials
  science,'' {\em Frontiers in Materials}, vol.~3, 2016.

\bibitem{Alghooneh2019}
A.~Alghooneh, S.~M.~A. Razavi, and S.~Kasapis, ``{Classification of
  hydrocolloids based on small amplitude oscillatory shear, large amplitude
  oscillatory shear, and textural properties.},'' {\em Journal of Texture
  Studies}, vol.~50, pp.~520--538, dec 2019.

\bibitem{Rudin2019}
C.~Rudin, ``{Stop explaining black box machine learning models for high stakes
  decisions and use interpretable models instead},'' {\em Nature Machine
  Intelligence}, vol.~1, no.~5, pp.~206--215, 2019.

\bibitem{Yegnanarayana2009}
B.~Yegnanarayana, {\em Artificial Neural Networks}.
\newblock PHI Learning Pvt. Ltd., 2009.

\bibitem{Molnar2020}
C.~Molnar, G.~Casalicchio, and B.~Bischl, ``Interpretable machine learning –
  a brief history, state-of-the-art and challenges,'' {\em Communications in
  Computer and Information Science}, p.~417–431, 2020.

\bibitem{Molnar2020book}
C.~Molnar, {\em Interpretable Machine Learning}.
\newblock 2~ed., 2022.

\bibitem{brenner2021}
M.~P. Brenner and P.~Koumoutsakos, ``Editorial: Machine learning and physical
  review fluids: An editorial perspective,'' {\em Phys. Rev. Fluids}, vol.~6,
  p.~070001, Jul 2021.

\bibitem{plazek1965}
D.~J. Plazek, ``Temperature dependence of the viscoelastic behavior of
  polystyrene,'' {\em The Journal of Physical Chemistry}, vol.~69,
  pp.~3480--3487, Oct 1965.

\bibitem{larsen2008}
T.~H. Larsen and E.~M. Furst, ``Microrheology of the liquid-solid transition
  during gelation,'' {\em Phys. Rev. Lett.}, vol.~100, p.~146001, Apr 2008.

\bibitem{dekker2018}
R.~I. Dekker, M.~Dinkgreve, H.~de~Cagny, D.~J. Koeze, B.~P. Tighe, and D.~Bonn,
  ``Scaling of flow curves: Comparison between experiments and simulations,''
  {\em Journal of Non-Newtonian Fluid Mechanics}, vol.~261, pp.~33--37, 2018.

\bibitem{caggioni2020}
M.~Caggioni, V.~Trappe, and P.~T. Spicer, ``Variations of the
  {H}erschel–{B}ulkley exponent reflecting contributions of the viscous
  continuous phase to the shear rate-dependent stress of soft glassy
  materials,'' {\em Journal of Rheology}, vol.~64, no.~2, pp.~413--422, 2020.

\bibitem{gupta2012}
R.~Gupta, B.~Baldewa, and Y.~M. Joshi, ``Time temperature superposition in soft
  glassy materials,'' {\em Soft Matter}, vol.~8, pp.~4171--4176, 2012.

\bibitem{struik1977}
L.~Struik, {\em Physical aging in amorphous polymers and other materials}.
\newblock PhD thesis, Delft University of Technology, Delft, NL, 11 1977.

\bibitem{lalwani2021}
S.~M. Lalwani, P.~Batys, M.~Sammalkorpi, and J.~L. Lutkenhaus, ``{Relaxation
  Times of Solid-like Polyelectrolyte Complexes of Varying pH and Water
  Content},'' {\em Macromolecules}, vol.~54, pp.~7765--7776, sep 2021.

\bibitem{markovitz1975}
H.~Markovitz, ``Superposition in rheology,'' {\em Journal of Polymer Science:
  Polymer Symposia}, vol.~50, no.~1, pp.~431--456, 1975.

\bibitem{Adolf1990}
D.~Adolf and J.~E. Martin, ``{Time-cure superposition during crosslinking},''
  {\em Macromolecules}, vol.~23, pp.~3700--3704, jul 1990.

\bibitem{Joshi2018}
Y.~M. Joshi and G.~Petekidis, ``{Yield stress fluids and ageing},'' {\em
  Rheologica Acta}, vol.~57, no.~6, pp.~521--549, 2018.

\bibitem{honerkamp1993}
J.~Honerkamp and J.~Weese, ``{A note on estimating mastercurves},'' {\em
  Rheologica Acta}, vol.~32, no.~1, pp.~57--64, 1993.

\bibitem{buttlar1998}
W.~G. Buttlar, R.~Roque, and B.~Reid, ``Automated procedure for generation of
  creep compliance master curve for asphalt mixtures,'' {\em Transportation
  Research Record}, vol.~1630, pp.~28--36, jan 1998.

\bibitem{sihn1999}
S.~Sihn and S.~W. Tsai, ``Automated shift for time-temperature superposition,''
  {\em Proceedings of the 12th International Comittee on Composite Materials},
  vol.~51, p.~47, 1999.

\bibitem{barbero2004}
E.~J. Barbero and K.~J. Ford, ``Equivalent time temperature model for physical
  aging and temperature effects on polymer creep and relaxation,'' {\em Journal
  of Engineering Materials and Technology}, vol.~126, pp.~413--419, nov 2004.

\bibitem{gergesova2011}
M.~Gergesova, B.~Zupančič, I.~Saprunov, and I.~Emri, ``The closed form
  t-{T}-{P} shifting ({CFS}) algorithm,'' {\em Journal of Rheology}, vol.~55,
  no.~1, pp.~1--16, 2011.

\bibitem{hermida1994}
{\'{E}}.~B. Hermida and F.~Povolo, ``Analytical-numerical procedure to
  determine if a set of experimental curves can be superimposed to form a
  master curve,'' {\em Polymer Journal}, vol.~26, no.~9, pp.~981--992, 1994.

\bibitem{naya2013}
S.~Naya, A.~Meneses, J.~Tarr{\'{i}}o-Saavedra, R.~Artiaga,
  J.~L{\'{o}}pez-Beceiro, and C.~Gracia-Fern{\'{a}}ndez, ``{New method for
  estimating shift factors in time–temperature superposition models},'' {\em
  Journal of Thermal Analysis and Calorimetry}, vol.~113, no.~2, pp.~453--460,
  2013.

\bibitem{cho2009}
K.-S. Cho, ``Geometric interpretation of time-temperature superposition,'' {\em
  Korea-Australia Rheology Journal}, vol.~21, no.~1, pp.~13--16, 2009.

\bibitem{maiti2016}
A.~Maiti, ``{A geometry-based approach to determining time-temperature
  superposition shifts in aging experiments},'' {\em Rheologica Acta}, vol.~55,
  no.~1, pp.~83--90, 2016.

\bibitem{Rouleau2013}
L.~Rouleau, J.-F. De{\"{u}}, A.~Legay, and F.~{Le Lay}, ``{Application of
  Kramers–Kronig relations to time–temperature superposition for
  viscoelastic materials},'' {\em Mechanics of Materials}, vol.~65, pp.~66--75,
  2013.

\bibitem{maiti2019}
A.~Maiti, ``{Second-order statistical bootstrap for the uncertainty
  quantification of time-temperature-superposition analysis},'' {\em Rheologica
  Acta}, vol.~58, no.~5, pp.~261--271, 2019.

\bibitem{matheron1963}
G.~Matheron, ``{Principles of geostatistics},'' {\em Economic Geology},
  vol.~58, pp.~1246--1266, dec 1963.

\bibitem{rasmussen2006}
C.~E. Rasmussen and C.~K.~I. Williams, {\em {Gaussian Processes for Machine
  Learning}}.
\newblock Cambridge, MA: MIT Press, 2006.

\bibitem{ewoldt2015}
R.~H. Ewoldt, M.~T. Johnston, and L.~M. Caretta, ``{Experimental Challenges of
  Shear Rheology: How to Avoid Bad Data},'' in {\em Complex Fluids in
  Biological Systems} (S.~E. Spagnolie, ed.), pp.~207--241, New York, NY:
  Springer New York, 2015.

\bibitem{singh2019}
P.~K. Singh, J.~M. Soulages, and R.~H. Ewoldt, ``{On fitting data for parameter
  estimates: residual weighting and data representation},'' {\em Rheologica
  Acta}, vol.~58, no.~6, pp.~341--359, 2019.

\bibitem{freund2015}
J.~B. Freund and R.~H. Ewoldt, ``Quantitative rheological model selection: Good
  fits versus credible models using {B}ayesian inference,'' {\em Journal of
  Rheology}, vol.~59, no.~3, pp.~667--701, 2015.

\bibitem{gortler2019}
J.~G{\"o}rtler, R.~Kehlbeck, and O.~Deussen, ``A visual exploration of
  {G}aussian processes,'' {\em Distill}, 2019.
\newblock https://distill.pub/2019/visual-exploration-gaussian-processes.

\bibitem{duvenaud2014}
D.~K. Duvenaud, {\em Automatic Model Construction with Gaussian Processes}.
\newblock PhD thesis, University of Cambridge, Cambridge, U.K., 6 2014.

\bibitem{Plazek1996}
D.~J. Plazek, ``1995 {B}ingham medal address: Oh, thermorheological simplicity,
  wherefore art thou?,'' {\em Journal of Rheology}, vol.~40, no.~6,
  pp.~987--1014, 1996.

\bibitem{Eggers2015}
J.~Eggers and M.~A. Fontelos, {\em {Singularities: Formation, Structure, and
  Propagation}}.
\newblock Cambridge: Cambridge University Press, 2015.

\bibitem{gelman2013}
A.~Gelman, J.~B. Carlin, H.~S. Stern, D.~B. Dunson, A.~Vehtari, and D.~B.
  Rubin, {\em {Bayesian Data Analysis}}.
\newblock Boca Raton, FL: CRC press, third~ed., 2013.

\bibitem{larsen2008kaj}
T.~Larsen, K.~Schultz, and E.~M. Furst, ``Hydrogel microrheology near the
  liquid-solid transition,'' {\em Korea-Australia Rheology Journal}, vol.~20,
  no.~3, pp.~165--173, 2008.

\bibitem{Hoerl1970}
A.~E. Hoerl and R.~W. Kennard, ``{Ridge Regression: Biased Estimation for
  Nonorthogonal Problems},'' {\em Technometrics}, vol.~12, pp.~55--67, feb
  1970.

\bibitem{zhang1993}
P.~Zhang, ``Model selection via multifold cross validation,'' {\em The Annals
  of Statistics}, vol.~21, pp.~299--313, mar 1993.

\bibitem{xu2004}
Q.-S. Xu, Y.-Z. Liang, and Y.-P. Du, ``Monte carlo cross-validation for
  selecting a model and estimating the prediction error in multivariate
  calibration,'' {\em Journal of Chemometrics}, vol.~18, no.~2, pp.~112--120,
  2004.

\bibitem{hastie2009}
T.~Hastie, R.~Tibshirani, and J.~Friedman, {\em The Elements of Statistical
  Learning: Data Mining, Inference, and Prediction}.
\newblock New York, NY: Springer Science \& Business Media, 2009.

\bibitem{silmore2019}
K.~S. Silmore, X.~Gong, M.~S. Strano, and J.~W. Swan, ``High-resolution
  nanoparticle sizing with maximum a posteriori nanoparticle tracking
  analysis,'' {\em ACS Nano}, vol.~13, pp.~3940--3952, apr 2019.

\bibitem{Suman2021}
K.~Suman, S.~Shanbhag, and Y.~M. Joshi, ``Phenomenological model of
  viscoelasticity for systems undergoing sol–gel transition,'' {\em Physics
  of Fluids}, vol.~33, no.~3, p.~033103, 2021.

\bibitem{Paredes2013}
J.~Paredes, M.~A.~J. Michels, and D.~Bonn, ``Rheology across the
  zero-temperature jamming transition,'' {\em Phys. Rev. Lett.}, vol.~111,
  p.~015701, Jul 2013.

\bibitem{Fielding2000}
S.~M. Fielding, P.~Sollich, and M.~E. Cates, ``Aging and rheology in soft
  materials,'' {\em Journal of Rheology}, vol.~44, no.~2, pp.~323--369, 2000.

\end{thebibliography}

\end{document}